\documentclass[journal=jpccck,manuscript=article]{achemso}
\setkeys{acs}{articletitle=true,etalmode=truncate}

\usepackage[version=3]{mhchem} 
\usepackage[T1]{fontenc}       
\usepackage{graphicx}
\usepackage{color}
\usepackage{float}
\usepackage[normalem]{ulem}

\author{Roberto Sant}
\affiliation{Univ. Grenoble Alpes, CEA, IRIG/DEPHY/MEM, 38000 Grenoble, France}
\alsoaffiliation{Univ. Grenoble Alpes, CNRS, Grenoble INP, Institut NEEL, 38000 Grenoble, France}
\alsoaffiliation{ESRF, The European Synchrotron, 38043 Grenoble, France}
\author{Simone Lisi} 
\affiliation{Univ. Grenoble Alpes, CNRS, Grenoble INP, Institut NEEL, 38000 Grenoble, France}
\author{Van Dung Nguyen}
\affiliation{Univ. Grenoble Alpes, CNRS, Grenoble INP, Institut NEEL, 38000 Grenoble, France}
\author{Estelle Mazaleyrat}
\affiliation{Univ. Grenoble Alpes, CEA, IRIG/DEPHY/PHELIQS, 38000 Grenoble, France}
\author{Ana Cristina G\'{o}mez Herrero}
\affiliation{Univ. Grenoble Alpes, CNRS, Grenoble INP, Institut NEEL, 38000 Grenoble, France}
\author{Olivier Geaymond}
\affiliation{Univ. Grenoble Alpes, CNRS, Grenoble INP, Institut NEEL, 38000 Grenoble, France}
\author{Val\'{e}rie Guisset}
\affiliation{Univ. Grenoble Alpes, CNRS, Grenoble INP, Institut NEEL, 38000 Grenoble, France}
\author{Philippe David}
\affiliation{Univ. Grenoble Alpes, CNRS, Grenoble INP, Institut NEEL, 38000 Grenoble, France}
\author{Alain Marty}
\affiliation{Univ. Grenoble Alpes, CEA, CNRS, Grenoble INP, IRIG/DEPHY/SpinTec, 38000 Grenoble, France}
\author{Matthieu Jamet}
\affiliation{Univ. Grenoble Alpes, CEA, CNRS, Grenoble INP, IRIG/DEPHY/SpinTec, 38000 Grenoble, France}
\author{Claude Chapelier}
\affiliation{Univ. Grenoble Alpes, CEA, IRIG, IRIG/DEPHY/PHELIQS, 38000 Grenoble, France}
\author{Laurence Magaud} 
\affiliation{Univ. Grenoble Alpes, CNRS, Grenoble INP, Institut NEEL, 38000 Grenoble, France}
\author{Yannick J. Dappe}
\affiliation{SPEC, CEA, CNRS, Universit\'{e} Paris-Saclay, CEA Saclay 91191 Gif-sur-Yvette Cedex, France}
\author{Marco Bianchi}
\affiliation{Department of Physics and Astronomy, Interdisciplinary Nanoscience Center (iNANO), Aarhus University, 8000 Aarhus C, Denmark}
\author{Philip Hofmann}
\affiliation{Department of Physics and Astronomy, Interdisciplinary Nanoscience Center (iNANO), Aarhus University, 8000 Aarhus C, Denmark}
\author{Gilles Renaud}
\affiliation{Univ. Grenoble Alpes, CEA, IRIG/DEPHY/MEM, 38000 Grenoble, France}
\author{Johann Coraux}
\affiliation{Univ. Grenoble Alpes, CNRS, Grenoble INP, Institut NEEL, 38000 Grenoble, France}
\email{johann.coraux@neel.cnrs.fr}

\title{Decoupling Molybdenum Disulfide from its Substrate by Cesium Intercalation}


\begin{document}


\begin{abstract}
Intercalation of alkali atoms within the lamellar transition metal dichalcogenides is a possible route toward a new generation of batteries. It is also a way to induce structural phase transitions authorizing the realization of optical and electrical switches in this class of materials. The process of intercalation has been mostly studied in three-dimensional dichalcogenide films. Here, we address the case of a single-layer of molybdenum disulfide (MoS$_2$), deposited on a gold substrate, and intercalated with cesium (Cs) in ultra-clean conditions (ultrahigh vacuum). We show that intercalation decouples MoS$_2$ from its substrate. We reveal electron transfer from Cs to MoS$_2$, relative changes in the energy of the valence band maxima, and electronic disorder induced by structural disorder in the intercalated Cs layer. Besides, we find an abnormal lattice expansion of MoS$_2$, which we relate to immediate vicinity of Cs. Intercalation is thermally activated, and so is the reverse process of de-intercalation. Our work opens the route to a microscopic understanding of a process of relevance in several possible future technologies, and shows a way to manipulate the properties of two-dimensional dichalcogenides by "under-cover" functionalization.
\end{abstract}


\section{Introduction}

The interest for transition metal dichalcogenide single-layers, initially spurred by the bright light emission found in molybdenum disulfide (MoS$_2$) \cite{Splendiani2010,Mak2010} and the achievement of electrostatic switching of electrical conduction in MoS$_2$,\cite{Radisavljevic2011} has revived activities devoted to the synthesis of these materials. Efforts to elaborate them with a structural quality similar to the one obtained in mechanically exfoliated samples with bottom-up approaches -- chemical vapour deposition, \cite{Zhan2012,Liu2012,Lee2012,Eichfeld2015} chalcogenation of metal surfaces,\cite{Kim2011,Orofeo2014,Wang2015} or molecular beam epitaxy, standard\cite{Ugeda2014} or reactive under H$_2$S atmosphere\cite{Helveg2000,Sorensen2014,Bana2018} -- are ongoing. Both chalcogenation and reactive molecular beam epitaxy usually require a metallic substrate. As-prepared samples are hence not suited to the study of some of the key properties of the material, \textit{e.g.} those related to excitons which become very short-lived due to the immediate vicinity of a metallic (substrate) charge reservoir, and electrical transport properties which are shunt by the conductive substrate. Besides, in the prototypical case of Au(111) as a substrate, MoS$_2$ does not retain the properties of the isolated material. Significant interaction between the electronic bands of MoS$_2$ and Au(111) was indeed detected\cite{Bruix2016} and the existence of a moir\'{e} pattern was found to induce a nanometer-scale modulation of this interaction.\cite{Krane2018}

One way to alter this interaction is to "lift" MoS$_2$ from its surface. Actually, such lifting occurs spontaneously, across regions spanning typically a nanometer, when MoS$_2$ overhangs on {\AA}ngstr\"{o}m-deep vacancy islands of the substrate.\cite{Krane2016} Effective lifting may be achieved using an alternative strategy, namely by intercalating a layer of a species decoupling MoS$_2$ from its substrate. This strategy, which allowed to obtain quasi-free-standing graphene\cite{Varykhalov2008,Riedl2009} (another two-dimensional material), has been explored recently with single-layer WS$_2$,\cite{Mahatha2018} but not with single-layer MoS$_2$ so far, to our knowledge.

On the contrary, intercalation of thicker transition metal dichalcogenides has been thoroughly investigated. Much like with graphite, a rich variety of systems with modulated structure in the direction perpendicular to the basal plane can be formed this way.\cite{Friend1987} Using layers of alkali atoms, molecules, or transition metals as intercalants, unique properties including superconductivity and (anti)ferromagnetism have been found.\cite{Friend1987} The ability to store (release) alkali atoms by intercalation (de-intercalation) also makes transition metal dichalcogenides possible electrode materials, both as cathode\cite{Murugan2006} and anode,\cite{Whittingham1976,Feng2009,Li2019} for Li-ion batteries. Electro-donor intercalants promote a structural phase transition from a semiconducting phase to a metallic one,\cite{Py1983,Wypych1992,Heising1999,Eda2011,Wang2013,Wang2014,Fan2015,Guo2015,Xiong2015,Ahmad2017} with potential applications in data storage and reconfigurable electrical circuitry.

Here, we report on the intercalation and de-intercalation of the alkali cesium (Cs) atoms. Unlike all works addressing in-solution intercalation of bulk-like transition metal dichalcogenide layers, the focus of our work is on single-layer MoS$_2$ flakes, prepared on Au(111), and (de)intercalated under ultrahigh vacuum conditions. We find that the process of intercalation is thermally activated, being completed after few tens of minutes at a temperature of 550~K. Above 850~K, deintercalation is efficient and completed within a few tens of minutes. Intercalated cesium forms a a Cs monolayer with an ill-ordered structure compatible with a $(\sqrt{3}\times\sqrt{3})R30^\circ$ reconstruction with respect to Au(111). We reveal electron transfer from Cs to MoS$_2$, modifications of the relative positions of the valence band maxima in MoS$_2$, and electronic disorder induced by structural disorder in the intercalated layer. Upon intercalation, MoS$_2$ is lifted, and adopts an unusually large lattice parameter. Our analysis combines scanning tunneling microscopy (STM), reflection high-energy electron diffraction (RHEED), grazing incidence X-ray diffraction (GIXRD), reflectivity (XRR), X-ray photoelectron spectroscopy (XPS), and angle-resolved photoemission spectroscopy (ARPES) all performed under ultrahigh vacuum, in some cases \textit{in operando} during intercalation. Further insights are brought by density functional theory (DFT) calculations.

\section{Methods}

Three ultrahigh vacuum systems were used for our experiments. A first one is coupled to the X-ray synchrotron beam delivered at the BM32 beamline of the ESRF. It has a base pressure of 3$\times$10$^{-10}$~mbar and is equipped with a quartz micro-balance and a RHEED apparatus. The second one, at Institut N\'{e}el (Grenoble), with a base pressure of 2$\times$10$^{-10}$~mbar, is part of a larger ultrahigh vacuum system comprising a STM, a RHEED apparatus, and a quartz microbalance. The samples were prepared in each system before being investigated by RHEED, STM, GIXRD, and XRR. Temperatures were measured with a pyrometer in both systems. Note that the pyrometers and the chamber configurations are different in the two systems, which implies a plausible variability ($\sim$50~K) in the measurements, and suggests caution when comparing these measurements. The third ultrahigh vacuum system is installed at the SGM-3 endstation\cite{Hoffmann2004} of the ASTRID2 synchrotron radiation source (Aarhus); it includes one chamber (base pressure 10$^{-10}$~mbar) devoted to ARPES and XPS measurements, and two other chambers (base pressures, 4$\times$10$^{-10}$~mbar) comprising a STM. All three ultrahigh vacuum systems are equipped with a Cs evaporator. MoS$_2$/Au(111) was prepared in Grenoble and transported in atmospheric conditions to Aarhus where it was degassed in ultrahigh vacuum at 500~K. There, temperatures were measured using a K type thermocouple fixed on the rear side of the Au (111) crystal. The cleanliness of the surface was confirmed with STM and photoelectron spectroscopy.

Single-crystals bought from Surface Preparation Laboratory and Mateck were prepared under ultrahigh vacuum by repeated cycles of room temperature sputtering with 0.8-1~keV Ar$^+$ ions and annealing to 900~K. Surface cleanliness was checked with STM imaging and RHEED, both revealing a well-developped herringbone reconstruction. Molybdenum was evaporated using a high-purity rod heated by electron-beam bombardment, at a rate of 0.02~monolayer/min in the ultrahigh vacuum chamber coupled to the X-ray beam and in the ultrahigh vacuum chamber coupled to the STM, respectively (one monolayer referring to the surface coverage of a single-layer MoS$_2$ on Au(111)). This rate was determined with both a quartz microbalance and STM. For introduction of H$_2$S in the ultrahigh vacuum chambers, we used an automatic injection system by VEGATEC that supplies H$_2$S via pneumatic valves (chamber coupled to the X-ray beam), and a leak-valve (chamber coupled to the STM). The latter system comprises large copper parts. Their surface was saturated by maintaining a pressure of 10$^{-6}$~mbar of H$_2$S for 30~min. Without this treatment, residual gas analysis revealed that H$_2$S was prominently decomposed before even reaching the sample surface, which prevented MoS$_2$ growth.

Cesium was deposited under ultrahigh vacuum by resistive heating of a high purity Cs dispenser (SAES Getter). Assessing the deposited Cs dose is not a straightforward task. Cesium dose measurements, using a quartz microbalance for instance, are seldom reported. Exceptions include low temperature measurements;\cite{Taborek1992} in contrast, our room temperature measurements were inconclusive. Alternative calibration methods were hence needed. In the ultrahigh vacuum system where STM imaging was performed, graphene growth on Ir(111) is operative. Cesium deposition in this system leads to its intercalation between graphene and Ir(111) in a sequence of well-crystallized phases readily detected with electron diffraction.\cite{Petrovic2013} At fixed current flowing the Cs dispenser (6.5~A) we hence determined the deposition time required for the onset of a $(\sqrt{3}\times\sqrt{3})R30^\circ$ Cs reconstruction relative to the graphene lattice, which corresponds to a well-defined Cs density. Using this calibration, we deposited Cs in two steps, each at room temperature followed by a 500~K annealing, with a total nominal Cs quantity equivalent to 0.7 Cs atoms per Au atom on the surface. In the ultrahigh vacuum system were XPS and ARPES data were acquired, a second calibration method was used. Cesium was deposited (5.7~A current) for a given time at room temperature onto Au(111), and the Cs density was assessed by comparing the area under the Cs peaks and Au surface 4$f$ components in XPS, assuming that Cs is in the form of a flat sub-monolayer deposit. With this calibration at hand, we then deposited Cs in two steps, each at room temperature followed by 550~K annealing, corresponding nominally to 0.3 and 0.6 Cs atoms per Au atom on the surface. In the third ultrahigh vacuum system, installed at the BM32 beamline, we could not use any of these two methods. We hence decided to deposit a large excess of Cs, using a high current flowing the Cs dispenser (7.2~A) and large deposition times, each of 30~min. There also we deposited Cs sequentially, in three steps, each at room temperature followed by a 500~K annealing.

Diffraction measurements were performed at European Synchrotron Radiation Facility using a $z$-axis diffractometer installed at the BM32 CRG/IF beamline and optimized for grazing incidence surface X-ray diffraction. The experimental energy was set at 11.8~keV, below the Au $L_\mathrm{III}$ absorption edge. The incident angle was set to 0.24$^\circ$, slightly below the critical angle for total external reflection, to enhance the signal from MoS$_2$ while minimizing the background. The diffraction signal was acquired with a Maxipix two-dimensional detector (256$\times$1288 pixels, each of size 55~$\mu$m).

The XRR spectra were processed with \texttt{PyRod} (home-made software for surface diffraction 2D data treatment). Data were integrated in reciprocal space and the profiles were extracted as a function of the modulus of the scattering vector perpendicular to the surface ($Q_\mathrm{\perp}$). Corrections to this integrated intensity were applied for beam polarization, surface active area and gaussian beam profile. Correction for beam refraction at interfaces were used to calculate $Q_\mathrm{\perp}$ inside the sample. The \texttt{ANA-ROD} software\cite{Vlieg2000} was used for modeling the surface structure of the samples and fitting the XRR data.

The STM data we show were acquired in Grenoble with an Omicron STM-1 apparatus, at room temperature under ultrahigh vacuum, in a dedicated chamber with a base pressure of 5$\times$10$^{-11}$~mbar. At the SGM-3 endstation, an Aarhus STM was used to check the sample quality after Cs deposition, which gave consistent results with those obtained in Grenoble.

ARPES and XPS measurements were performed at the SGM-3 end station of the ASTRID2 synchrotron radiation source.\cite{Hoffmann2004} The ARPES and XPS data were collected at room temperature at photon energies of 49~eV and 130~eV respectively. The energy and angular resolutions were better than 20~meV and 0.2$^\circ$ respectively.

Structural relaxations as well as electronic structure determination have been performed using DFT. The DFT localized-orbital molecular-dynamics code as implemented in \texttt{Fireball}\cite{Lewis2011,Jelinek2005,Sankey1989} has been used to optimize the MoS$_2$/Au(111) and the MoS$_2$/Cs/Au(111) structures or to determine their corresponding electronic properties. Standard previously used basis sets have been considered for Mo, S and Au,\cite{Gonzalez2016} and an $sp$ basis set with cut-off radii of 6.8 (in atomic units), has been considered for Cs. A hexagonal slab of 7$\times$7 Au atoms with five layers in the $xy$ plane, and a topmost 6$\times$6 MoS$_2$ have been used to model the MoS$_2$/Au(111) interface. This configuration has been optimized until the forces become less than 0.1~eV/\AA. The bottom two Au layers were fixed to simulate the Au bulk for the MoS$_2$/Au(111) interface. The MoS$_2$/Cs/Au(111) system, with a $(\sqrt{3}\times\sqrt{3})R30^\circ$ Cs reconstruction on Au(111) requires a large supercell to be treated. In this systems, the atomic positions were fixed to the average values derived from the analysis of the XRR data. We have used a set of 32 $k$-points in the plane for self-consistency and density of states calculations.

\section{Results and Discussion}

\begin{figure*}[!hbt]
\centering
\includegraphics[width=14.76cm]{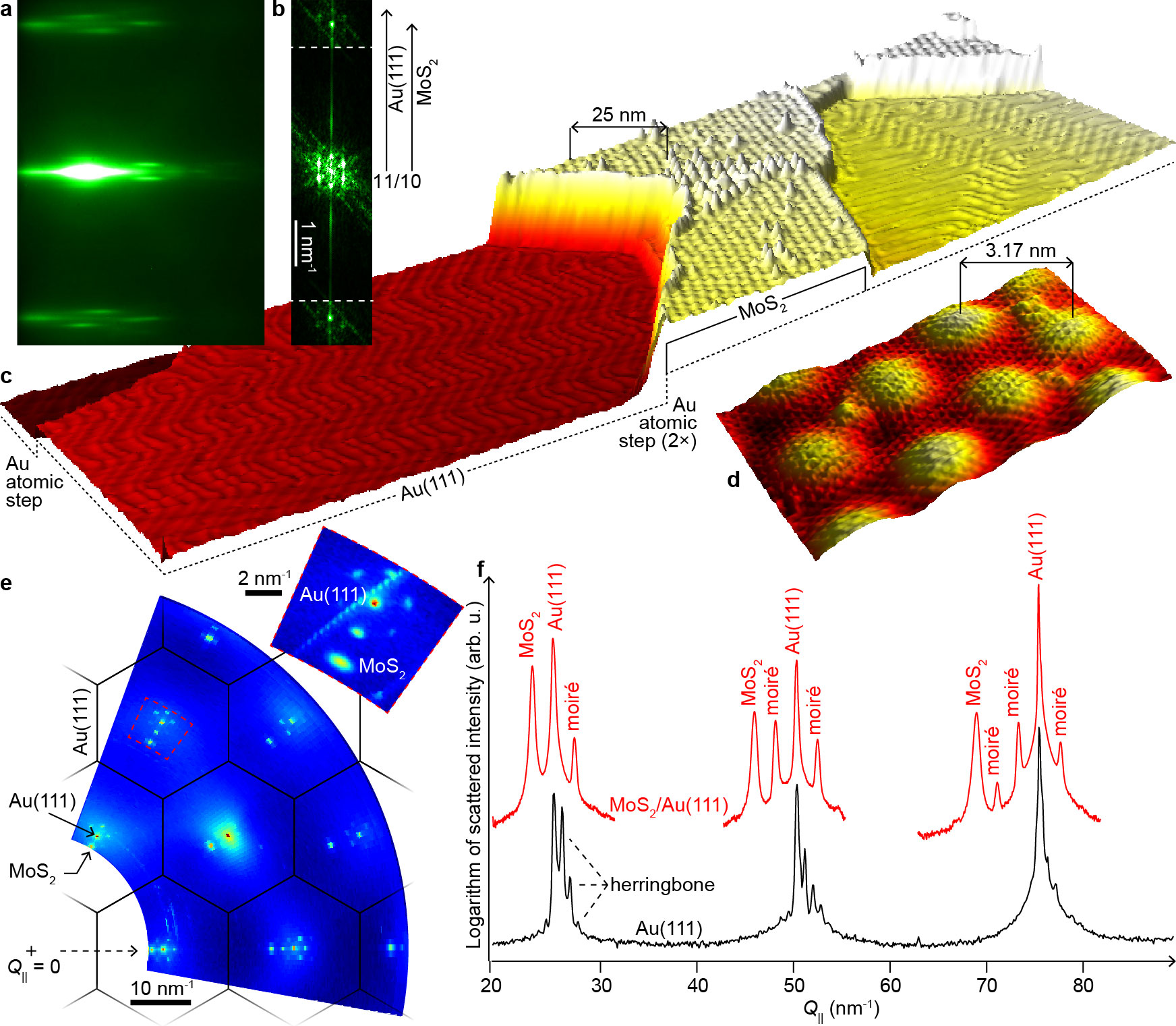}
\caption{Single layer MoS$_2$ islands prepared on Au(111). (\textit{a}) RHEED diffractogram (17~keV) along the $[1\bar{1}0]$ azimuth. On either side of the most intense central specular streak, MoS$_2$ and moir\'{e} streaks are observed, the latter ones corresponding to the shortest vertical distances on the pattern. (\textit{b}) Fast Fourier transform of an atomically resolved STM image (the contrast and brightness below and above the dotted line, at the vicinity of the MoS$_2$ spots, have been enhanced). The reciprocal space vectors corresponding to the Au(111) and MoS$_2$ lattices are highlighted with two arrows, whose length ratio is 11/10. (\textit{c}) Three-dimensional view of a STM topograph (2~nA, -2~V) of single-layer MoS$_2$ islands formed on Au(111), with the herringbone reconstruction of the latter visible. Defects in the moir\'{e} appear as high bumps. (\textit{d}) Three-dimensional view of an atomically-resolved STM topograph (0.81~nA, -1.93~V) of the moir\'{e} lattice between Au(111) and MoS$_2$. (\textit{e}) In-plane cut of the reciprocal lattice, as measured with X-rays. Top right: higher-resolution measurement of the area marked with a dotted frame. (\textit{f}) Radial scan of the X-ray scattered intensity in the direction shown with a dotted arrow in \textit{e}, as a function of the modulus of the in-plane scattering vector $Q_\mathrm{\parallel}$ for Au(111) with and without single-layer MoS$_2$ islands on top.}
\label{fig1}
\end{figure*}

Single-layer MoS$_2$ flakes were grown following the procedure described by Gr{\o}nborg \textit{et al.}\cite{Gronborg2015} In short, the clean Au(111) surface was exposed to a partial pressure of H$_2$S ($P_\mathrm{H_2S}$) introduced in the ultrahigh vacuum systems, then molybdenum was deposited on the surface in presence of H$_2$S, and the sample was annealed to 900~K without the Mo atomic beam but still in presence of H$_2$S. This sequence was repeated several times to adjust the surface coverage with MoS$_2$. In the two ultrahigh vacuum chambers where MoS$_2$ was grown, we used $P_\mathrm{H_2S}$ = 10$^{-5}$ and 10$^{-6}$~mbar respectively. Figure~\ref{fig1}a-d shows the typical diffraction patterns and morphology of the surface after growth. The MoS$_2$ flakes exhibit straight edges,\cite{Bollinger2000} have an extension of the order of several 10~nm; in between the flakes the herringbone reconstruction of the bare Au(111) is visible. The fraction of the surface covered with MoS$_2$ was 70$\pm$15\%, 28$\pm$5\%, and 25$\pm$15\% (as determined from the nominal amount of Mo deposited on the surface, or with STM whenever possible) for the samples studied with GIXRD/XRR, STM, and XPS/ARPES, respectively. Our STM observations are consistent with those in previous reports.\cite{Sorensen2014,Gronborg2015}

\textbf{Structure of MoS$_2$ single-layer islands on Au(111).} A pronounced pattern, with 3.17$\pm$0.01~nm periodicity, is visible on the MoS$_2$ flakes presented in Figure~\ref{fig1}c,d. This pattern arises from the lattice mismatch between MoS$_2$ and the substrate, and is described with an analogy to the optical moir\'{e} effect.\cite{Sorensen2014} Careful analysis of atomically-resolved STM images and their Fourier transform (Figure~\ref{fig1}b) allows to determine the size of the moir\'{e} unit cell. For the example shown in Figure~\ref{fig1}, we find that the highest symmetry Au(111) and MoS$_2$ crystallographic directions are precisely aligned, and that the unit cell corresponds to the coincidence of 10 MoS$_2$ unit cells onto 11 Au(111) unit cells (10$\times$11), consistent with a recent report.\cite{Bana2018}

This moir\'{e} unit cell is varying from one MoS$_2$ island to another. The average reciprocal space lattice vector associated to the moir\'{e} is directly inferred from RHEED (Figure~\ref{fig1}a) and GIXRD (Figure~\ref{fig1}e,f) data by measuring the relative positions of the moir\'{e} or MoS$_2$ peaks with respect to the Au(111) peaks, to be 1.884 $\pm$ 0.001~nm$^{-1}$.

The full-width at half maximum of the MoS$_2$ peaks in a radial direction (Figure~\ref{fig1}f) increases from 0.32 $\pm$ 0.05~nm$^{-1}$ to 0.51 $\pm$ 0.05~nm$^{-1}$ from first to third order. This corresponds to a domain size of about 20~nm and a distribution of in-plane lattice parameter of typically 0.6\%.\cite{Guinier1994} Strikingly, the domain size is here smaller than the value of several 10~nm corresponding to the flake size that we determined by visual inspection of STM images. This difference simply shows that the flakes are not single-crystal, and actually consist each of (smaller) single-crystal grains. As discussed in the Supporting Information (SI), we indeed frequently observe linear defects within the flakes, at the boundary between laterally-shifted domains within the flake (see Figure~S1). Our interpretation is that at each of the several steps of the MoS$_2$ cyclic preparation, new MoS$_2$ islands nucleate, grow, and coalesce with pre-existing ones -- no lattice re-organisation occurs that would eliminate the linear defect (so-called out-of-phase grain boundaries) to yield large single-crystal flakes.

To model the structure of MoS$_2$ and learn more about the out-of-plane structure of the material, we use DFT calculations taking van der Waals interactions into account. For that purpose, we choose an unsheared moir\'{e} unit cell with 6 MoS$_2$ units matching 7 Au(111) surface units. This (6$\times$7) coincidence lattice is not the (10$\times$11) observed experimentally. Our choice is however legitimate since in our DFT calculations, the lattice structure optimization are performed at 0~K (while the measurements are performed at 300~K), a temperature at which the lattice parameters of bulk Au(111) (0.2883~nm) and MoS$_2$ (0.3293~nm) are in a different ratio (1.14, \textit{i.e.} close to 7/6) than those measured for bulk compounds at room temperature (0.2884~nm and 0.3167~nm, ratio = 1.10 = 11/10, see Ref.~\citenum{ElMahalawy1976}). The DFT commensurate structure is a realistic approximant of the experimental unit cell and we expect to capture the physics of interaction. This structure is also computationally more demanding than the (1$\times$1) and $(\sqrt{13}\times\sqrt{13})R13.9^\circ$ commensurate approximants used so far.\cite{Bruix2016}

\begin{figure}[!hbt]
\centering
\includegraphics[width=8.25cm]{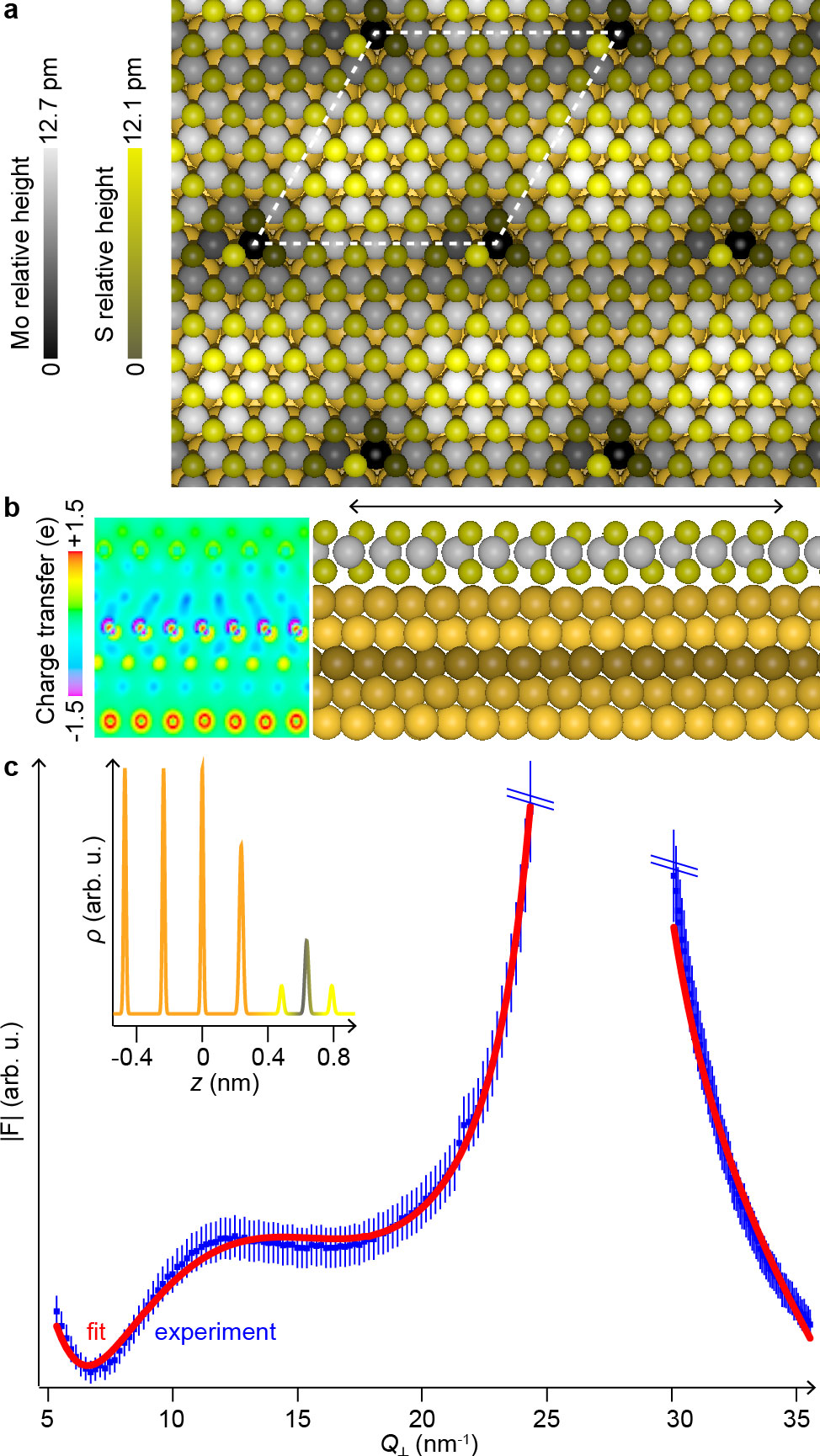}
\caption{Nature of the bonding for MoS$_2$ onto Au(111). (\textit{a}) Top-view of the optimized geometry found with DFT calculations. The Mo (S) atoms are sketched with gray (yellow) balls, whose shade codes the variation of height. The rhombus highlights the commensurate (6$\times$7) moir\'{e} unit cell employed for the calculations. (\textit{b}) Cross-section of the charge transferred between atoms, along an edge of the rhombus, and of the atomic structure, along the long diagonal of the rhombus (of length indicated with a double arrow), calculated with DFT. The different layers of the Au(111) substrate are coloured according to the sequence of ABC planes in a $fcc$ stack. (\textit{c}) Experimental structure factor $F$ modulus (XRR, blue dots) as function of the modulus of the out-of-plane scattering vector $Q_\mathrm{\perp}$, and best fit to the experimental data (red curve). Inset: Electronic density profile versus out-of-plane coordinate ($z$), corresponding to the best fit.}
\label{fig2}
\end{figure}

Figure~\ref{fig2}a shows a top-view sof the minimum-energy structure of MoS$_2$ on Au(111) that has been optimised with the DFT calculations. Periodic lattice distorsions are found in both MoS$_2$ and Au(111). The in-plane projection of the distorsions of the topmost Au(111) layer and of the Mo layer are better visualised in Figure~S2, where they have been amplified. They range from typically 1 to 8~pm (Au), and 1 to 4~pm (Mo). In another epitaxial two-dimensional material, graphene, smaller ($\sim$1~pm) and larger ($\sim$20~pm) distorsions have been reported on Ir(111) (Ref.~\citenum{Blanc2012}) and Ru(0001) (Ref.~\citenum{Martoccia2010}) substrates. In the former system no or very weak moir\'{e} peaks were observed while they were clearly observed in the latter.\cite{Martoccia2008} Our observation of intense moir\'{e} diffraction peaks in MoS$_2$/Au(111), and the fact that among these peaks, those located closer to the Au(111) diffraction peaks are those with higher intensity (Figure~\ref{fig1}f), are consistent with significant distorsions in the Au lattice.

Regarding the out-of-plane structure (Figure~\ref{fig2}b), according to the DFT calculations the average distance between the topmost Au plane and the closest S plane, $d_\mathrm{Au-S}$, amounts 0.252~nm, while the average distance between the Mo and top (bottom) S planes is $d_\mathrm{Mo-S}$ = 0.153~nm (0.156~nm). The $d_\mathrm{Au-S}$ interlayer value is smaller than the 0.312~nm interlayer distance value found in our DFT calculations for an infinite multilayer of 2H-MoS$_2$, where the interplanar interactions are of van der Waals type. We note that the $d_\mathrm{Mo-S}$ values, on the contrary, are very similar to the one obtained with multilayer of 2H-MoS$_2$. The values of $d_\mathrm{Au-S}$ and $d_\mathrm{Mo-S}$ are modulated along the moir\'{e} pattern, by only few 10~pm and few 1~pm respectively. This suggests that the apparent height modulations of typically 100~pm observed with STM in relation with the moir\'{e} pattern, are essentially of electronic nature.

\begin{figure*}[!hbt]
\centering
\includegraphics[width=7.85cm]{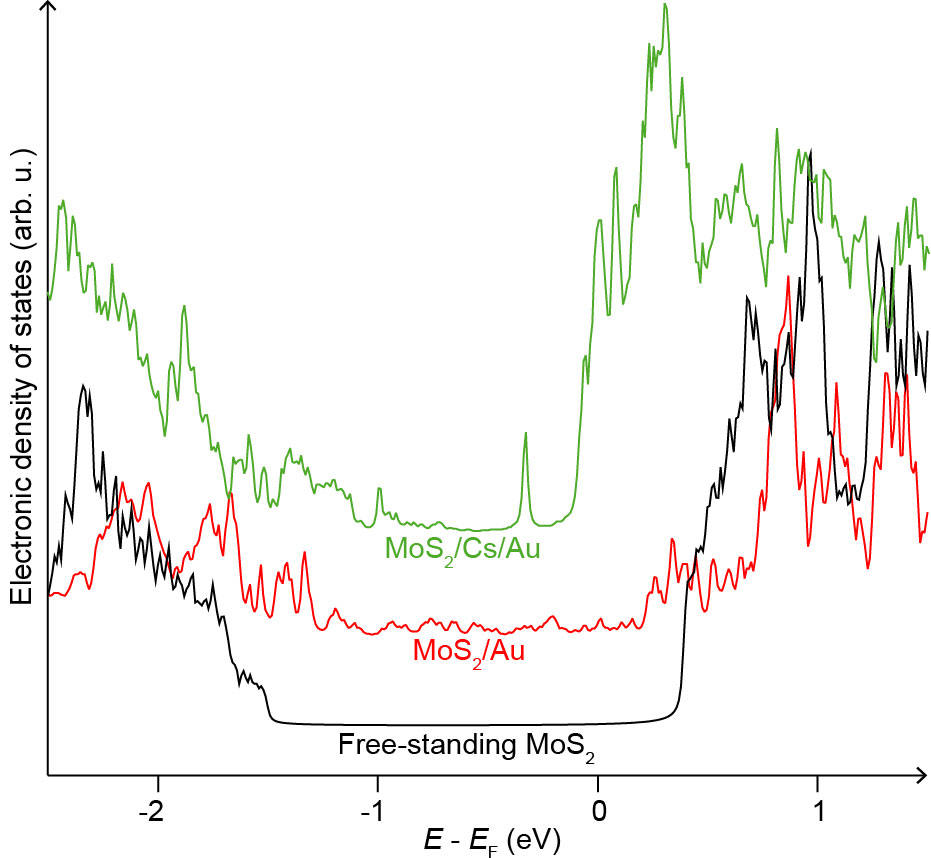}
\caption{Electronic density of states derived from DFT calculations for free-standing MoS$_2$, MoS$_2$ on Au(111) in a (6$\times$7) coincidence lattice, and MoS$_2$ on Au(111) with an intercalated $(\sqrt{3}\times\sqrt{3})R30^\circ$ Cs layer.}
\label{figDOS}
\end{figure*}

The short $d_\mathrm{Au-S}$ values shows that strong Au-S bonds exist in the system, and this is confirmed by the DFT analysis of the charge transfer at the interface between MoS$_2$ and Au(111) (Figure~\ref{fig2}b). Within the unit cell, a spatial modulation of the charge held by the S atoms in the bottom layer and by the Au top layers is observed. This is a signature of a spatially varying interaction between MoS$_2$ and Au(111), which is expected when hybridization between the electronic orbitals of MoS$_2$ and Au(111) occurs locally within the moir\'{e}, where the atoms belonging to each material are positioned properly. The bottom S layers has a deficit of electrons (blue shades in Figure~\ref{fig2}b), while the Mo and top S layers rather have an excess of electrons (green to light-green shades in Figure~\ref{fig2}b). This is consistent with the electronic density of states, also calculated with DFT, which shows that the bottom of the conduction band is down-shifted by about 160~meV compared to the case of free-standing MoS$_2$ (Figure~\ref{figDOS}, see also discussion in the SI and Figure~S3). Overall the DFT analysis indicates that the MoS$_2$ single layer on Au(111) is slight electron-doped.

We now confront the DFT analysis to the experimental analysis of the out-of-plane structure via XRR. The modulus of the Fourier transform of the total electronic density of the system along the out-of-plane direction (Figure ~\ref{fig2}c) is obtained by taking the square root of the XRR data. In between the (000) and (111) crystallographic reflections of Au(111) (modulus of the out-of-plane scattering vector $Q_\mathrm{\perp}$, 0~nm$^{-1}$ and 26.68~nm$^{-1}$), we observe a bump in the reflectivity. Qualitatively, the local reflectivity minima correspond to destructive interferences between the X-ray waves scattered by the Au(111) lattice and by the MoS$_2$ layer, perpendicular to the surface. The distance between the minima is related to the distances between the layers (it decreases when the interlayer distance increases, and \textit{vice versa}).

We performed a more quantitative analysis by calculating the structure factor of an atomic model of MoS$_2$/Au(111), and refining the values of the free structural parameters to fit the calculation to the experimental data, using the \texttt{ANA-ROD} code.\cite{Vlieg2000} In short (see Figure~S4a and Table~S1 for details), we assume two regions that scatter incoherently, one with the bare Au(111) surface, the other with MoS$_2$/Au(111), consistent with the partial coverage of the surface with MoS$_2$. We disregard the above-discussed (weak) periodic distorsion, to limit the number of free parameters and thus to ensure a reliable fit. The topmost Au (111) atomic plane is let free to move in the $z$ direction, to account for surface relaxation. The above-defined $d_\mathrm{Au-S}$ and $d_\mathrm{Mo-S}$ distances are also free parameters for the fit. Finally, the roughness of the surface is modeled within the so-called $\beta$ model,\cite{Robinson1986} with $\beta$ as another free parameter. We obtain a very good fit to the experimental data (Figure~\ref{fig2}c) for the electronic density profile shown in the inset of Figure~\ref{fig2}c, that yields refined values of $d_\mathrm{Au-S}$ = 0.243 $\pm$ 0.025~nm and $d_\mathrm{Mo-S}$ = 0.153 $\pm$ 0.022~nm. These values are in very good agreement with the outcome of the DFT calculations.

\begin{figure*}[!hbt]
\centering
\includegraphics[width=12.56cm]{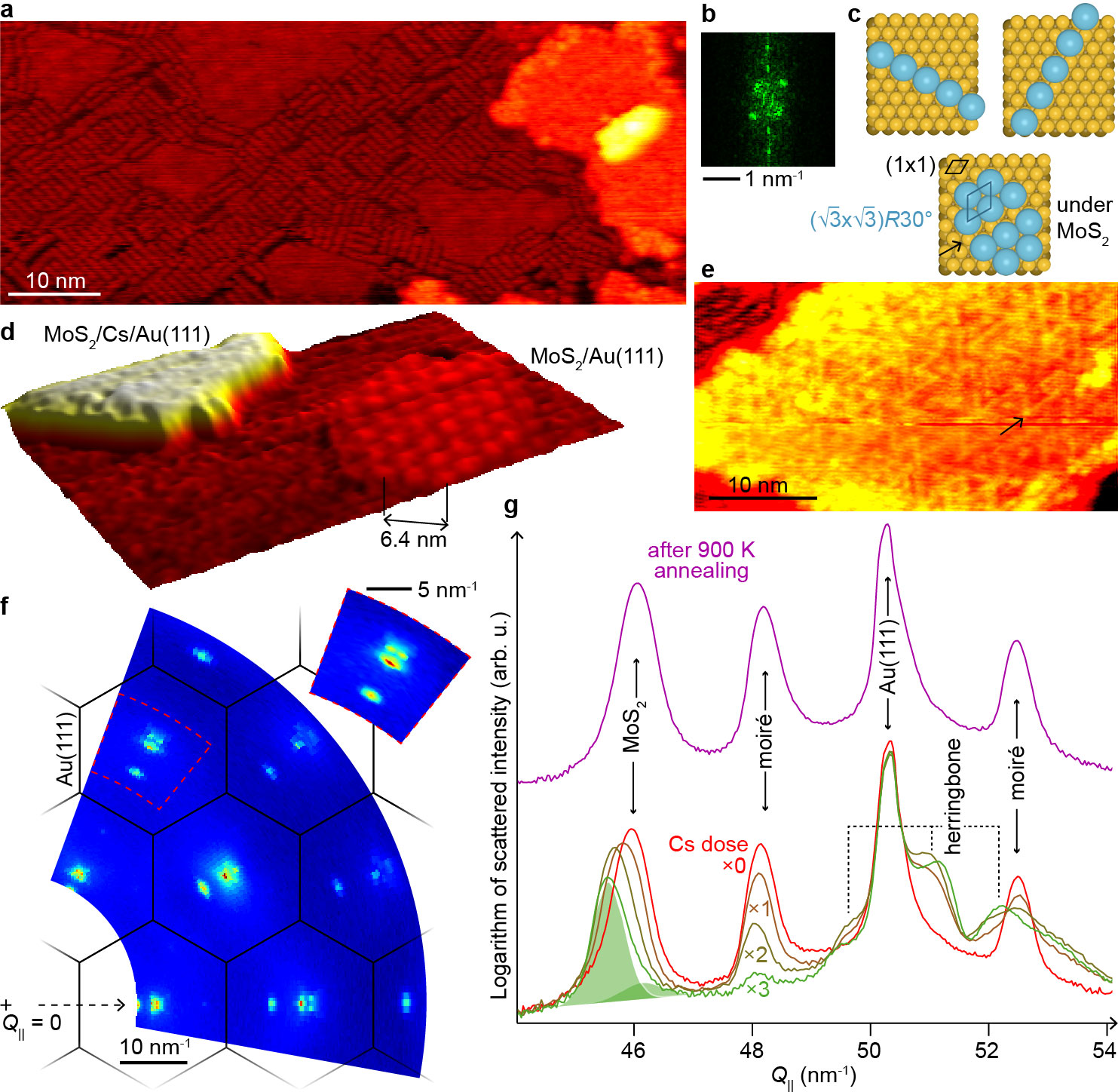}
\caption{Structural changes in MoS$_2$ upon Cs intercalation. (\textit{a}) STM view (0.2~nA, -2~V) revealing Cs nanosticks on Au(111) after deposition (nominally 0.7 Cs atoms per Au surface atom) at room temperature and annealing to 500~K. (\textit{b}) Fast Fourier transform of a region with nanosticks. (\textit{c}) Cartoons of the possible structures for the Cs nanosticks and of the $(\sqrt{3}\times\sqrt{3})R30^\circ$ Cs phase intercalated under MoS$_2$. (\textit{d}) Three-dimensional view of a STM topograph (0.65~nA, -0.5~V)  showing MoS$_2$ islands with and without intercalated Cs. (\textit{e}) Close-up STM view (0.2~nA, -2~V) of a MoS$_2$ island revealing a pattern of lines. (\textit{f}) In-plane  reciprocal map measured with X-rays after the deposition of three excess Cs doses and 550~K annealing. Top-right: high-resolution measurement in the area marked with a red-dashed frame. (\textit{g}) Radial scan of the X-ray scattered intensity versus the in-plane momentum transfer $Q_\mathrm{\parallel}$. The bottom four scans have been measured each after the additional deposition of an excess Cs dose and subsequent annealing at 550~K, while the top scan has been measured after the final Cs deposit/annealing followed by an annealing at 900~K. For the largest dose the two components (green areas) used to fit the MoS$_2$ peak are shown.}
\label{fig3}
\end{figure*}

\textbf{Cesium deposition and intercalation of Cs underneath MoS$_2$.} STM was performed after the nominal deposition of 0.7 Cs per Au surface atom and 500~K annealing. On MoS$_2$-free regions Cs atoms form a pattern of $\sim$5~nm-long nanosticks, some bunching across several 1~nm to several 10~nm, and having different orientations on the surface, also detected in the Fourier transform of the images (Figure~\ref{fig3}a,b). The strong structural disorder evident on the STM image explains the absence of a Cs-related diffraction signature at room temperature in RHEED and GIXRD. Considering the observed local periodicity, we propose two possible atomic structures for the Cs nanosticks (Figure~\ref{fig3}c).

We observe two kinds of MoS$_2$-covered regions (Figure~\ref{fig3}a,d,e). Part of the MoS$_2$ has a similar height as the Cs-covered Au(111) surface, and exhibits a moir\'{e} pattern. The majority ($\sim$85\%) of MoS$_2$ islands however has a higher apparent height and exhibits no moir\'{e} pattern. A reasonable interpretation is that the latter islands are intercalated with a Cs layer, while the former islands are not. This intercalated layer is not directly accessible to STM measurements. We observe a nanoscale pattern of lines oriented along the three highest-symmetry directions of Au(111) and MoS$_2$ (120$^\circ$ orientations). This pattern may be related to that observed in another intercalated two-dimensional-material-on-metal system, graphene/Bi/Ir(111).\cite{Warmuth2016} There, the pattern was interpreted as a network of dislocations in the intercalant's lattice, the lines corresponding to the boundaries between intercalated domains being shifted by a fraction of the lattice vector(s) of the intercalant's lattice (Figure~\ref{fig3}c).

The disappearance of the moir\'{e} pattern observed in STM is corroborated by GIXRD measurements. After three cycles of deposition of a large Cs excess followed by 550~K annealing, the reciprocal space lattice of the sample shows no discernable moir\'{e} signals (compare Figures~\ref{fig3}f,\ref{fig1}e). In fact, the reduction of the moir\'{e} signals rather occurs after than before each 550~K annealing (Figure~\ref{fig3}g). The process is hence thermally-activated, indicative of kinetically limitations.

Concomitantly to the vanishing of the moir\'{e} signal, the GIXRD data reveal that the diffraction signal associated to MoS$_2$ progressively shows two components, the one at lower scattering vector modulus corresponding to an expansion of the lattice (Figure~\ref{fig3}g). The latter component may be assigned to intercalated islands, while the other corresponds to pristine, still-unintercalated MoS$_2$ islands on Au. This suggests that intercalation proceeds sequentially, island by island. In other words, the limiting kinetic step in the intercalation process corresponds to the opening of an intercalation channel, for instance a point defect or the unbinding of (part of) the flake edges from the substrate. Once this channel is opened, mass transport underneath the flake is presumably very efficient at the several-10~nm-scale considered here. At the end of the three deposition+annealing cycles, close to 100\% of the islands are intercalated (note that the intensity scale in Figure~\ref{fig3}g is logarithmic).

The GIXRD data show a decrease in intensity and an angular broadening of the diffraction peaks as the Cs dose increases (compare Figures~\ref{fig3}f,\ref{fig1}e), which both point to increased disorder in the form of in-plane strains and mosaic spread. Conversely, we observe a resurgence of the diffraction signals of the Au reconstruction (Figure~\ref{fig3}g), suggesting that the Au-MoS$_2$ interaction has been suppressed.

\textbf{Strain induced by intercalation.} In the absence of a significant hybridization between MoS$_2$ and Au orbitals, the MoS$_2$ is no more strongly pinned on the substrate lattice. Consistent with this view, the in-plane lattice parameter of MoS$_2$ is found to change (the MoS$_2$ diffraction peak is shifted to lower momentum transfer values, Figure~\ref{fig3}g), \textit{i.e.} the MoS$_2$ lattice is no longer strained by its substrate. More precisely, a 0.9\% lattice expansion is observed at 300~K compared to the value found in bulk MoS$_2$.\cite{ElMahalawy1976}

Several effects may explain this expansion. First, MoS$_2$ is grown at 900~K, a temperature at which the ratio of lattice parameters for MoS$_2$ and Au(111) is expected to be precisely 1.10,\cite{ElMahalawy1976,Huang2014,Nix1941} \textit{i.e.} precisely the 11/10 ratio determined experimentally after the sample is cooled down to room temperature. This is yet another indication that prior to intercalation MoS$_2$ is strongly bond on its substrate, with an epitaxial relationship that is unchanged between growth temperature and room temperature. In the opposite extreme case, if during cool down MoS$_2$ were free to compress according to its own natural thermal compression (and not that of the Au(111) substrate, which is slightly larger), the lattice parameter of MoS$_2$ would be slightly larger, by $\sim$0.2\%.\cite{ElMahalawy1976,Huang2014,Nix1941} This is the maximum expansion we expect due to the suppression of the strong bonding between MoS$_2$ and Au upon intercalation. This effect hence only accounts for a small fraction of the observed 0.9\% expansion.

What is the origin of the remaining $\sim$0.7\% expansion? A structural phase transition (1H to 1T or 1T') is expected upon electron doping,\cite{Brumme2015} and Cs, a well-known electro-donor species, might indeed donate the required amount of charges to MoS$_2$. The 1T phase is not expected to have a significantly different lattice constant (a marginally shorter lattice constant has even been predicted\cite{Duerloo2014}) while the 1T' phase should.\cite{Tan2018} However our diffraction measurements do not detect a doubling of the MoS$_2$ unit cell that would be expected for this phase. There is another reason why an increasing amount of alkali atoms in the vicinity of MoS$_2$ could actually lead to an increased lattice constant. In the related system of potassium inserted in between MoS$_2$ layers, DFT calculations predicted a significant lattice expansion. There, the role of the increased charge density within the Mo-S bonds was invoked.\cite{Andersen2012} Our own DFT calculations with free-standing MoS$_2$ (this is a rough, yet reasonable description of MoS$_2$ decoupled from Au(111) by intercalation) predict that adding about 0.1 electrons per MoS$_2$ unit cell (to mimik the charge transfer associated to intercalation) yields a lattice expansion by typically several 0.1\%.

\begin{figure}[!hbt]
\centering
\includegraphics[width=8.25cm]{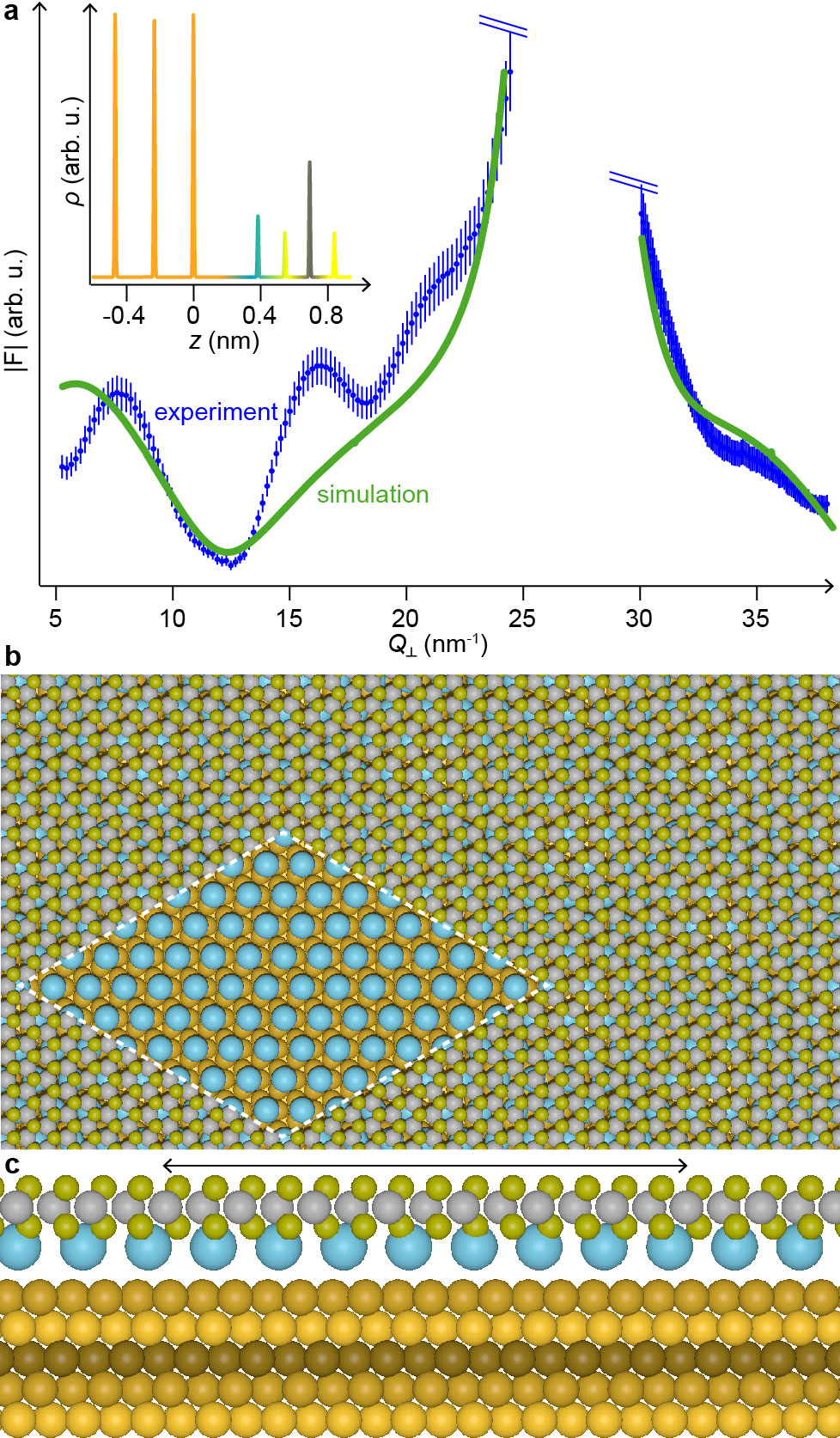}
\caption{Nature of the structure and bonding after Cs intercalation. (\textit{a}) Experimental structure factor $F$ modulus (XRR, blue dots) as function of $Q_\mathrm{\perp}$ after three deposits of an excess Cs + 550~K annealing, and best simulation to the experimental data (green curve). Inset: Corresponding electronic density profile along $z$. (\textit{b}) Top-view of the geometry optimized from the XRR analysis and used for DFT calculations. The Mo (S) atoms are sketched with gray (yellow) balls, whose shade codes the variation of height. The Cs atoms are sketched with cyan balls. The rhombus highlights the commensurate (6$\times$7) $(\sqrt{30}\times\sqrt{30})R30^\circ$ moir\'{e} unit cell. Within the area of the rhombus, we do not show the MoS$_2$ atoms to display the underlying $(\sqrt{30}\times\sqrt{30})R30^\circ$ Cs layer. (\textit{c}) Cross-section, along the long diagonal of the rhombus, of the atomic structure, along the long diagonal of the rhombus, deduced from the analysis of the XRR data. The Au(111) layers are coloured according to the $fcc$ sequence of ABC planes.}
\label{fig4}
\end{figure}

We again examine the structure of the sample perpendicular to the surface. While for pristine MoS$_2$/Au(111), the XRR spectrum exhibited essentially one intensity rebound (Figure~\ref{fig2}c), after Cs deposition and annealing at least four rebounds ($\sim$ 8, 16, 22, 35~nm$^{-1}$) are observed (Figure~\ref{fig4}a). The distance between the extrema decreasing with increasing interplanar distances, this new observation is consistent with an increased $d_\mathrm{Au-S}$ distance, compared to the case of MoS$_2$/Au(111). This is precisely what is expected if a Cs layer is intercalated between MoS$_2$ and Au(111). To test this interpretation, we have adjusted $d_\mathrm{Au-S}$ (and the distance $d_\mathrm{Au-Cs}$ between the Cs layer and the Au topmost plane; see SI for further details). Given the large atomic radius of Cs, we expect a low-density Cs phase. Our STM analysis directly confirms that on MoS$_2$-free regions (Figure~\ref{fig3}a-c), the shortest Cs-Cs distances are indeed large, matching the second nearest neighbour Au-Au distance on the surface, \textit{i.e.} 0.499~nm. Two-dimensional materials tend to alter the organisation of intercalant as shown in the case of graphene,\cite{Martinez2016} hence an even lower density in the intercalated Cs layer, as \textit{e.g.} in a $(2\times2)$ phase, cannot be excluded \textit{a priori}. However our simulations agree less with the XRR data in the case of such a low-density phase, suggesting that a denser phase, for instance a $(\sqrt{3}\times\sqrt{3})R30^\circ$ reconstruction, provides a more realistic description of the system. The best match of our model to the experimental data is obtained for $d_\mathrm{Au-S}$ = 0.551~nm and $d_\mathrm{Au-Cs}$ = 0.389~nm (see Table~S2 for details). The $d_\mathrm{Au-S}$ value is increased by about 0.308~nm compared to the case without intercalated Cs. The increase is close, but 0.07~nm less than that observed in Cs-intercalated multilayer MoS$_2$.\cite{Somoano1973}

Figure~\ref{fig4}a obviously shows that this model is too simple to faithfully reproduce \textit{all} the features in the experimental XRR data. In particular, between 15~nm$^{-1}$ and 25~nm$^{-1}$, the model only accounts for the baseline of the experimental spectrum, and does not produce the two marked intensity rebounds. As discussed more into details in the SI (see Figure~S5 and Table~S3 for another simulation with a more advanced model), what is not captured by our simple model is the multilayer thickness of the Cs layer on bare Au(111) (unlike under MoS$_2$, where it is intercalated as a single-layer). Multilayer Cs is expected there because we chose to deposit a large excess of Cs on the surface for the samples characterized with XRR (and GIXRD), due to the lack of reliable calibration of the Cs deposition rate in the corresponding ultrahigh vacuum chamber (see \textit{Methods}). On MoS$_2$, long-ranged Cs surface diffusion and Cs desorption are instead expected already at room temperature to result in (\textit{i}) low-surface-density Cs clusters, and (\textit{ii}) mass transport of Cs out of the MoS$_2$ surface, towards MoS$_2$-free regions and vacuum. 550~K annealing will further promote this tendency, and strongly reduce the Cs cluster density on MoS$_2$.

\textbf{Thermal de-intercalation.} As already mentionned, increasing the temperature promotes intercalation of Cs underneath the MoS$_2$ flakes, which points to a kinetic barrier to intercalation (\textit{e.g.} for passing through defects and/or for creating defects later acting as intercalation pathways). To further improve the efficiency of intercalation it is tempting to further increase temperature. Above about 800~K however, another key process is activated: the moir\'{e} signal re-appears in GIXRD, the MoS$_2$ peak shifts back to its initial position (Figure~\ref{fig3}g), and the XRR spectrum strongly resembles that of as-deposited MoS$_2$/Au(111). In fact, after 900~K annealing, the reciprocal lattice of the sample is very similar to that of the pristine sample, \textit{i.e.} Cs has been de-intercalated. How is this occurring? An XPS analysis (see SI) reveals that a fraction of Cs adatoms penetrates the Au crystal already at room temperature, even more so at 550~K (Figure~S6) and at 850~K. In addition to this process, at 850~K Cs atoms may have a non-negligible probability to cross the energy barrier for diffusing outside the MoS$_2$ islands, and to subsequently desorb to vacuum.

\begin{figure*}[!hbt]
\centering
\includegraphics[width=14.41cm]{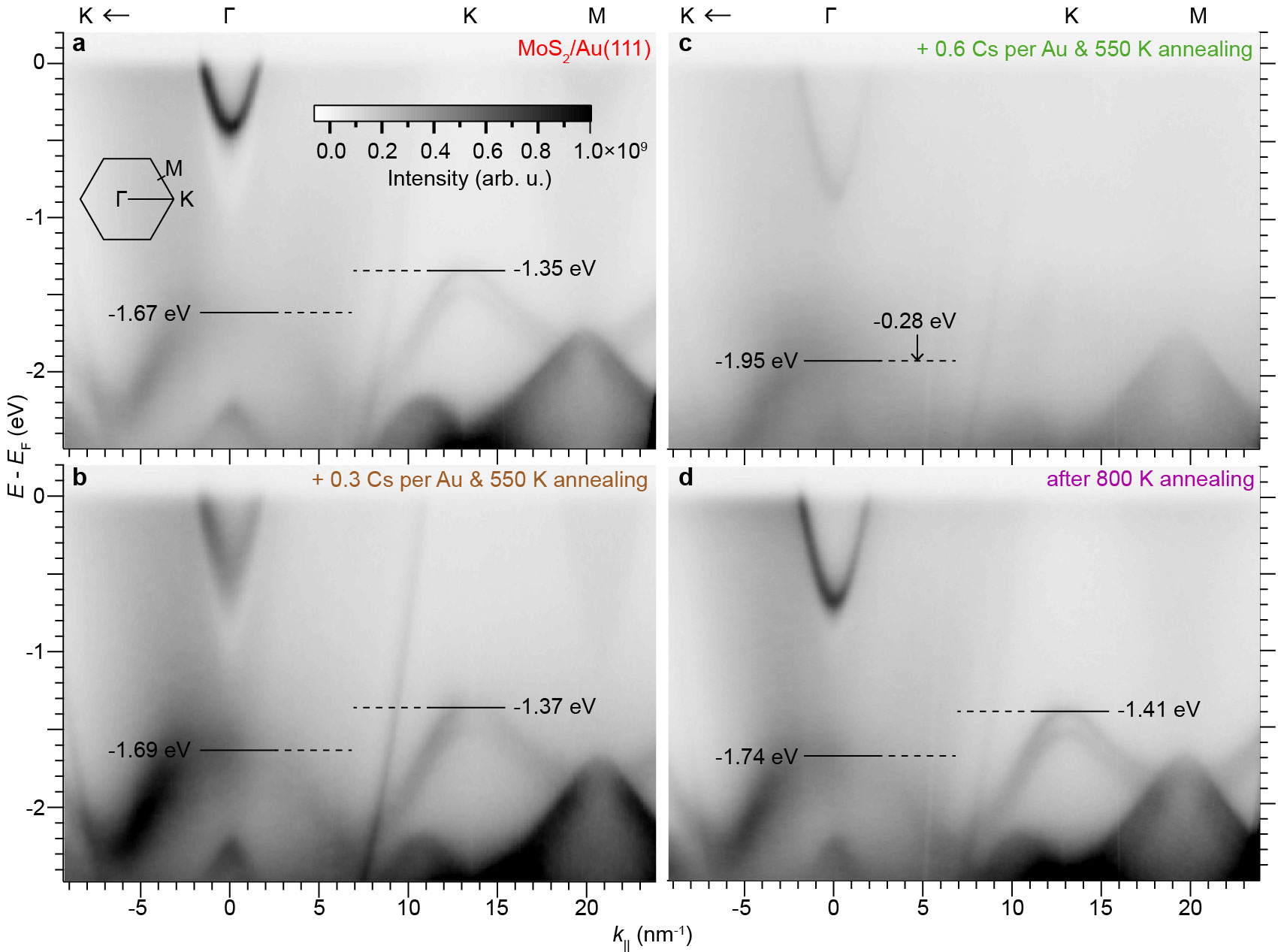}
\caption{Electronic modifications upon intercalation. Photoemission intensity along high-symmetry reciprocal space directions (M$\Gamma$KM, see inset), with a 49~eV photon energy, for (\textit{a}) pristine MoS$_2$/Au(111), and (\textit{b,c}) the same sample exposed to an increasing Cs dose +0.3 and +0.6 Cs atoms per Au surface atom (followed by 550~K annealing), and (\textit{d}) eventually annealed to 850~K. The black lines mark the positions of the valence band extrema before/after Cs intercalation.}
\label{fig5}
\end{figure*}

\textbf{Electronic effects of Cs intercalation.} We expect that intercalation of Cs has strong influence on the electronic properties of MoS$_2$. We start with experimental measurements, with ARPES, of energy-momentum cuts in the electronic band structure along K$\Gamma$KM in reciprocal space, first with the case of as-grown MoS$_2$/Au(111). The valence band of MoS$_2$ (whose maxima are highlighted with horizontal lines in Figure~\ref{fig5}), with a characteristic 130~meV spin-orbit splitting at K point, is clearly seen (Figure~\ref{fig5}a). Its maxima at $\Gamma$ and K points lie 1.67~eV and 1.35~eV respectively below the Fermi level, while the conduction band minimum is above the Fermi level, consistent with a previous report.\cite{Bruix2016}

After the deposition of nominally 0.3 Cs atom per surface Au atom, and thermally-induced intercalation the ARPES energy-momentum cuts (Figure~\ref{fig5}b) and energy distribution curves (EDCs, Figure~S7) show that the MoS$_2$ valence band maximum at the $\Gamma$ and K points shift down, by 20~meV. No additional electronic band is observed that would correspond to Cs. Tripling the Cs dose yields stronger changes in the electronic band structure of MoS$_2$ (Figures~\ref{fig5}c,S7): the MoS$_2$ valence band maximum at $\Gamma$ point further shifts down, by as much as 280~meV. The photoemission signal corresponding to the MoS$_2$ valence band at the vicinity of the K point becomes weaker and broader.Inspection of the EDCs (Figure~S7) is required to detect a down-shift, of about 100~meV.

The Au(111) surface state at the $\Gamma$ point, close to the Fermi level, is observed before and all along the Cs deposition/annealing procedure (Figure~\ref{fig5}). The signature of this surface state is naturally related to the occurence of the Au(111) herringbone reconstruction observed in STM in between MoS$_2$ islands (Figure~\ref{fig1}c) and with GIXRD after Cs deposition/annealing (Figure~\ref{fig3}g). This indicates that Cs is weakly adsorbed on Au, hence not altering the sensitive Au(111) reconstruction/surface state. We note that as the Cs dose increases, the surface state shifts down in binding energy (Figure~\ref{fig5}a-c). Shifts of this kind have been observed for Na adsorbed on Cu(111),\cite{Lindgren1980} and have been explained by a polarization-induced filling of the surface state.\cite{Forster2006}

These observations are reminiscent of a previous study that explored charge transfers induced by potassium (K) atoms in MoS$_2$/Au(111).\cite{Miwa2015} In the two studies, the most prominent effects seem to be a non-rigid down-shift of the electronic band structure, with different magnitude at $\Gamma$ and K points. The shifts are of the same order of magnitude with potassium and cesium, suggesting charge transfers (the Fermi level is changed) in the same range. The origin of the shifts can be qualitatively rationalised by inspecting the projection of the electronic states on the different orbitals in the system, which has been calculated by DFT for a (1$\times$1) approximate model for MoS$_2$ on Au(111).\cite{Bruix2016} The spin-orbit-split MoS$_2$ valence band close to K point is expected to be primarily of Mo $d_\mathrm{x^2-y^2}+d_\mathrm{xy}$ character and the fact that it is not significantly shifted upon intercalation suggests that it is not related to a possible hybridization with the substrate's electronic band (the hybridization would be strongly affected by intercalation), which seems reasonable for these in-plane MoS$_2$ orbitals. In the energy range explored in Figure~\ref{fig5}, at $\Gamma$ point the stronger contribution to the valence band stems from Mo $d_\mathrm{z^2}+d_\mathrm{yz}+d_\mathrm{xz}$ out-of-plane orbitals, and these bands are indeed expected to be involved in the hybridization and charge transfers with the substrate or the alkali atoms, consistent with our observations (Figure~\ref{fig5}).

The broadening of the valence band at K point after Cs intercalation points to a significant disorder in the system. This is consistent with our STM observations of a disordered nanoscaled pattern for Cs in this case (Figure~\ref{fig3}c,e). Strikingly, annealing the sample to 850~K allows to recover a well-defined valence band at K point (compare Figure~\ref{fig5}a,d), consistent with the deintercalation process also evident in GIXRD (Figure~\ref{fig3}g) and XRR, which we interpret as a consequence of Cs diffusion into bulk Au or Cs desorption from the surface. This is also an indication that the source of electronic disorder was indeed extrinsic to the MoS$_2$, \textit{i.e.} due to the intercalant, and not intrinsic to MoS$_2$, \textit{i.e.} due to the creation of defects in the dichalcogenide single-layer.

Our ARPES observation can be rationalised in the framework of a simplified DFT calculation scheme. Minimising the structure of the MoS$_2$/Cs$(\sqrt{3}\times\sqrt{3})R30^\circ$/Au(111) system would require to consider a large commensurate calculation supercell comprising about 1,100 atoms (in a $(\sqrt{3}\times\sqrt{3})R30^\circ$, and not a (1$\times$1), moir\'{e} unit cell), among which high-$Z$ number ones. This is computationally prohibitive, and we prefer, instead, to analyse the electronic density of states for the values of $d_\mathrm{Au-Cs}$ = 0.389~nm and $d_\mathrm{Au-S}$ = 0.551~nm (see structural model represented in Figures~\ref{fig4}b,c) which we derived from the analysis of the XRR data. Keeping in mind the limitations of the latter analysis, we also considered another set of $d_\mathrm{Au-Cs}$ and $d_\mathrm{Au-S}$ values, for which the Cs layer is further from (closer to) the MoS$_2$ layer (Au(111) surface), see SI for details.

Figure~\ref{figDOS} compares the electronic density of states in the presence of the intercalated layer ($d_\mathrm{Au-Cs}$ = 0.389~nm, $d_\mathrm{Au-S}$ = 0.551~nm) to the case of MoS$2$/Au(111) and the case of isolated MoS$_2$. The bottom of the conduction band is further down-shifted, by about 320~meV, relative to the case of MoS$_2$/Au(111), which is indicative of a strong electron doping of MoS$_2$, consistent with our ARPES observations. This tendency is robust and is also predicted for different $d_\mathrm{Au-Cs}$ and $d_\mathrm{Au-S}$ values (Figure~S3). The level of electronic doping seems comparable to the one that we expect for an isolated MoS$_2$ layer, with three Cs atoms adsorbed per $(10\times10)$ MoS$_2$ unit cells (Figure~S3).

Except for one marked peak (-330~meV) that we ascribe to an interfacial Au/Cs state, within about 600~meV below the bottom of the conduction band, the DFT calculations predict few in-gap states than for MoS$_2$/Au(111). Below this energy range more in-gap states are predicted, at least for $d_\mathrm{Au-Cs}$ = 0.389~nm and $d_\mathrm{Au-S}$ = 0.551~nm. Their occurrence is however highly dependent on the distance between the bottom S layer and the Cs layer, as shown for a calculation performed with another set of $d_\mathrm{Au-Cs}$ and $d_\mathrm{Au-S}$ values (Figure~S3).

\section{Conclusions}

We showed that starting from MoS$_2$ flakes strongly coupled to the Au(111) substrate, with a short spacing and a substantial nanorippling, Cs intercalation lifts the flakes and flatens them. The process is thermally activated. De-intercalation is also thermally activated, but at higher temperatures. We found that Cs, which is intercalated in the form of a atomic layer whose density is consistent with a $(\sqrt{3}\times\sqrt{3})R30^\circ$ reconstruction on Au(111), substantially dopes MoS$_2$ with electrons, and that this doping is a possible origin for a $\sim$1\% expansion of the atomic lattice of MoS$_2$ parallel to the surface. The interaction with the Cs layer is associated with relative changes in the energy of the valence band maxima and to electronic disorder presumably related to the structural disorder in Cs.

Our work opens new perspectives towards the manipulation of two-dimensional transition metal dichalcogenides. Similarly to in-solution strategies,\cite{Eda2011} intercalation could be exploited on MoS$_2$/Au(111) to facilitate the exfoliation of nanoscale flakes\cite{Helveg2000} or full layers.\cite{Bana2018} Demonstrating further control on electronic or hole doping of MoS$_2$ with intercalated electro-donor or electro-acceptor species is another exciting goal. A number of species, to be intercalated with varying doses, are relevant here, among the vast catalog of alkali atoms, transition metals, and molecules. Finally, as extensively demonstrated with bulk compounds in the past, intercalation opens new doors to achieve a variety of two-dimensional phases, structural ones, magnetic ones, and even superconducting ones.

\begin{acknowledgement}

R.S. acknowledges financial support by Nanosciences Foundation. This work was supported by the 2DTransformers project under the OH-RISQUE program (ANR-14-OHRI-0004), J2D (ANR-15-CE24-0017) and ORGANI'SO (ANR-15-CE09-0017) projects of Agence Nationale de la Recherche (ANR), by the French state funds Equipex ANR-11-EQPX-0010, ANR-10-LABX-51-01 (Labex LANEF of "Programme d'Investissements d'Avenir"), ANR-15-IDEX-02, and by the R\'{e}gion Rh\^{o}ne Alpes (ARC6 program) and the Labex LANEF. The was also supported by VILLUM FONDEN via the Centre of Excellence for Dirac Materials (Grant No. 11744). We thank Tao Zhou for his kind support.

\end{acknowledgement}

\begin{suppinfo}

Supporting information comprises additional STM data showing defects in MoS$_2$, the results of DFT calculations on deformations in MoS$_2$ and its substrate, energy distribution ARPES curves, and a DFT analysis of the the influence of an intercalated AuCs alloy on the electronic properties of MoS$_2$.

\end{suppinfo}


\begin{mcitethebibliography}{64}
\providecommand*\natexlab[1]{#1}
\providecommand*\mciteSetBstSublistMode[1]{}
\providecommand*\mciteSetBstMaxWidthForm[2]{}
\providecommand*\mciteBstWouldAddEndPuncttrue
  {\def\EndOfBibitem{\unskip.}}
\providecommand*\mciteBstWouldAddEndPunctfalse
  {\let\EndOfBibitem\relax}
\providecommand*\mciteSetBstMidEndSepPunct[3]{}
\providecommand*\mciteSetBstSublistLabelBeginEnd[3]{}
\providecommand*\EndOfBibitem{}
\mciteSetBstSublistMode{f}
\mciteSetBstMaxWidthForm{subitem}{(\alph{mcitesubitemcount})}
\mciteSetBstSublistLabelBeginEnd
  {\mcitemaxwidthsubitemform\space}
  {\relax}
  {\relax}

\bibitem[Splendiani \latin{et~al.}(2010)Splendiani, Sun, Zhang, Li, Kim, Chim,
  Galli, and Wang]{Splendiani2010}
Splendiani,~A.; Sun,~L.; Zhang,~Y.; Li,~T.; Kim,~J.; Chim,~C.-Y.; Galli,~G.;
  Wang,~F. Emerging Photoluminescence in Monolayer MoS$_2$. \emph{Nano Lett.}
  \textbf{2010}, \emph{10}, 1271--1275\relax
\mciteBstWouldAddEndPuncttrue
\mciteSetBstMidEndSepPunct{\mcitedefaultmidpunct}
{\mcitedefaultendpunct}{\mcitedefaultseppunct}\relax
\EndOfBibitem
\bibitem[Mak \latin{et~al.}(2010)Mak, Lee, Hone, Shan, and Heinz]{Mak2010}
Mak,~K.~F.; Lee,~C.; Hone,~J.; Shan,~J.; Heinz,~T.~F. Atomically Thin MoS$_2$:
  A New Direct-Gap Semiconductor. \emph{Phys. Rev. Lett.} \textbf{2010},
  \emph{105}, 136805\relax
\mciteBstWouldAddEndPuncttrue
\mciteSetBstMidEndSepPunct{\mcitedefaultmidpunct}
{\mcitedefaultendpunct}{\mcitedefaultseppunct}\relax
\EndOfBibitem
\bibitem[Radisavljevic \latin{et~al.}(2011)Radisavljevic, Radenovic, Brivio,
  Giacometti, and Kis]{Radisavljevic2011}
Radisavljevic,~B.; Radenovic,~A.; Brivio,~J.; Giacometti,~i.~V.; Kis,~A.
  Single-Layer MoS$_2$ Transistors. \emph{Nat. Nanotech.} \textbf{2011},
  \emph{6}, 147\relax
\mciteBstWouldAddEndPuncttrue
\mciteSetBstMidEndSepPunct{\mcitedefaultmidpunct}
{\mcitedefaultendpunct}{\mcitedefaultseppunct}\relax
\EndOfBibitem
\bibitem[Zhan \latin{et~al.}(2012)Zhan, Liu, Najmaei, Ajayan, and
  Lou]{Zhan2012}
Zhan,~Y.; Liu,~Z.; Najmaei,~S.; Ajayan,~P.~M.; Lou,~J. Large-Area Vapor-Phase
  Growth and Characterization of MoS$_2$ Atomic Layers on a SiO$_2$ Substrate.
  \emph{Small} \textbf{2012}, \emph{8}, 966--971\relax
\mciteBstWouldAddEndPuncttrue
\mciteSetBstMidEndSepPunct{\mcitedefaultmidpunct}
{\mcitedefaultendpunct}{\mcitedefaultseppunct}\relax
\EndOfBibitem
\bibitem[Liu \latin{et~al.}(2012)Liu, Zhang, Lee, Lin, Chang, Su, Chang, Li,
  Shi, Zhang, Lai, and Li]{Liu2012}
Liu,~K.-K.; Zhang,~W.; Lee,~Y.-H.; Lin,~Y.-C.; Chang,~M.-T.; Su,~C.-Y.;
  Chang,~C.-S.; Li,~H.; Shi,~Y.; Zhang,~H. \latin{et~al.}  Growth of Large-Area
  and Highly Crystalline MoS$_2$ Thin Layers on Insulating substrates.
  \emph{Nano Lett.} \textbf{2012}, \emph{12}, 1538--1544\relax
\mciteBstWouldAddEndPuncttrue
\mciteSetBstMidEndSepPunct{\mcitedefaultmidpunct}
{\mcitedefaultendpunct}{\mcitedefaultseppunct}\relax
\EndOfBibitem
\bibitem[Lee \latin{et~al.}(2012)Lee, Zhang, Zhang, Chang, Lin, Chang, Yu,
  Wang, Chang, Li, and Lin]{Lee2012}
Lee,~Y.-H.; Zhang,~X.-Q.; Zhang,~W.; Chang,~M.-T.; Lin,~C.-T.; Chang,~K.-D.;
  Yu,~Y.-C.; Wang,~J. T.-W.; Chang,~C.-S.; Li,~L.-J. \latin{et~al.}  Synthesis
  of Large-Area MoS$_2$ Atomic Layers with Chemical Vapor Deposition.
  \emph{Adv. Mater.} \textbf{2012}, \emph{24}, 2320--2325\relax
\mciteBstWouldAddEndPuncttrue
\mciteSetBstMidEndSepPunct{\mcitedefaultmidpunct}
{\mcitedefaultendpunct}{\mcitedefaultseppunct}\relax
\EndOfBibitem
\bibitem[Eichfeld \latin{et~al.}(2015)Eichfeld, Hossain, Lin, Piasecki, Kupp,
  Birdwell, Burke, Lu, Peng, Li, Azcati, McDonnel, Wallace, Kim, Mayer, Rewing,
  and Robinson]{Eichfeld2015}
Eichfeld,~S.~M.; Hossain,~L.; Lin,~Y.-C.; Piasecki,~A.~F.; Kupp,~B.;
  Birdwell,~A.~G.; Burke,~R.~A.; Lu,~N.; Peng,~X.; Li,~J. \latin{et~al.}
  Highly Scalable, Atomically Thin WSe$_2$ Grown via Metal-Organic Chemical
  Vapor Deposition. \emph{ACS Nano} \textbf{2015}, \emph{9}, 2080--2087\relax
\mciteBstWouldAddEndPuncttrue
\mciteSetBstMidEndSepPunct{\mcitedefaultmidpunct}
{\mcitedefaultendpunct}{\mcitedefaultseppunct}\relax
\EndOfBibitem
\bibitem[Kim \latin{et~al.}(2011)Kim, Sun, Lu, Cheng, Zhu, Le, Rahman, and
  Bartels]{Kim2011}
Kim,~D.; Sun,~D.; Lu,~W.; Cheng,~Z.; Zhu,~Y.; Le,~D.; Rahman,~T.~S.;
  Bartels,~L. Toward the Growth of an Aligned Single-Layer MoS$_2$ Film.
  \emph{Langmuir} \textbf{2011}, \emph{27}, 11650--11653\relax
\mciteBstWouldAddEndPuncttrue
\mciteSetBstMidEndSepPunct{\mcitedefaultmidpunct}
{\mcitedefaultendpunct}{\mcitedefaultseppunct}\relax
\EndOfBibitem
\bibitem[Orofeo \latin{et~al.}(2014)Orofeo, Suzuki, Sekine, and
  Hibino]{Orofeo2014}
Orofeo,~C.~M.; Suzuki,~S.; Sekine,~Y.; Hibino,~H. Scalable Synthesis of
  Layer-Controlled WS$_2$ and MoS$_2$ Sheets by Sulfurization of Thin Metal
  Films. \emph{Appl. Phys. Lett.} \textbf{2014}, \emph{105}, 083112\relax
\mciteBstWouldAddEndPuncttrue
\mciteSetBstMidEndSepPunct{\mcitedefaultmidpunct}
{\mcitedefaultendpunct}{\mcitedefaultseppunct}\relax
\EndOfBibitem
\bibitem[Wang \latin{et~al.}(2015)Wang, Li, Yao, Song, Sun, Pan, Ren, Li,
  Okunishi, Wang, Wang, Shao, Zhang, Yang, Schwier, Iwasawa, Shimada,
  Taniguchi, Cheng, Zhou, Du, Pennycook, Pantelides, and Gao]{Wang2015}
Wang,~Y.; Li,~L.; Yao,~W.; Song,~S.; Sun,~J.; Pan,~J.; Ren,~X.; Li,~C.;
  Okunishi,~E.; Wang,~Y.-Q. \latin{et~al.}  Monolayer PtSe$_2$, a New
  Semiconducting Transition-Metal-Dichalcogenide, Epitaxially Grown by Direct
  Selenization of Pt. \emph{Nano Lett.} \textbf{2015}, \emph{15},
  4013--4018\relax
\mciteBstWouldAddEndPuncttrue
\mciteSetBstMidEndSepPunct{\mcitedefaultmidpunct}
{\mcitedefaultendpunct}{\mcitedefaultseppunct}\relax
\EndOfBibitem
\bibitem[Ugeda \latin{et~al.}(2014)Ugeda, Bradley, Shi, Felipe, Zhang, Qiu,
  Ruan, Mo, Hussain, Shen, Wang, Louie, and Crommie]{Ugeda2014}
Ugeda,~M.~M.; Bradley,~A.~J.; Shi,~S.-F.; Felipe,~H.; Zhang,~Y.; Qiu,~D.~Y.;
  Ruan,~W.; Mo,~S.-K.; Hussain,~Z.; Shen,~Z.-X. \latin{et~al.}  Giant Bandgap
  Renormalization and Excitonic Effects in a Monolayer Transition Metal
  Dichalcogenide Semiconductor. \emph{Nat. Mater.} \textbf{2014}, \emph{13},
  1091--1095\relax
\mciteBstWouldAddEndPuncttrue
\mciteSetBstMidEndSepPunct{\mcitedefaultmidpunct}
{\mcitedefaultendpunct}{\mcitedefaultseppunct}\relax
\EndOfBibitem
\bibitem[Helveg \latin{et~al.}(2000)Helveg, Lauritsen, L{\ae}gsgaard,
  Stensgaard, N{\o}rskov, Clausen, Tops{\o}e, and Besenbacher]{Helveg2000}
Helveg,~S.; Lauritsen,~J.~V.; L{\ae}gsgaard,~E.; Stensgaard,~I.;
  N{\o}rskov,~J.~K.; Clausen,~B.; Tops{\o}e,~H.; Besenbacher,~F. Atomic-Scale
  Structure of Single-Layer MoS$_2$ Nanoclusters. \emph{Phys. Rev. Lett.}
  \textbf{2000}, \emph{84}, 951\relax
\mciteBstWouldAddEndPuncttrue
\mciteSetBstMidEndSepPunct{\mcitedefaultmidpunct}
{\mcitedefaultendpunct}{\mcitedefaultseppunct}\relax
\EndOfBibitem
\bibitem[S{\o}rensen \latin{et~al.}(2014)S{\o}rensen, F\"{u}chtbauer, Tuxen,
  Walton, and Lauritsen]{Sorensen2014}
S{\o}rensen,~S.~G.; F\"{u}chtbauer,~H.~G.; Tuxen,~A.~K.; Walton,~A.~S.;
  Lauritsen,~J.~V. Structure and Electronic Properties of In Situ Synthesized
  Single-Layer MoS$_2$ on a Gold Surface. \emph{ACS Nano} \textbf{2014},
  \emph{8}, 6788--6796\relax
\mciteBstWouldAddEndPuncttrue
\mciteSetBstMidEndSepPunct{\mcitedefaultmidpunct}
{\mcitedefaultendpunct}{\mcitedefaultseppunct}\relax
\EndOfBibitem
\bibitem[Bana \latin{et~al.}(2018)Bana, Travaglia, Bignardi, Lacovig, Sanders,
  Dendzik, Michiardi, Bianchi, Lizzit, Presel, Angelis, Apostol, Das, Fujii,
  Vobornik, Larciprete, Baraldi, Hofmann, and Lizzit]{Bana2018}
Bana,~H.; Travaglia,~E.; Bignardi,~L.; Lacovig,~P.; Sanders,~C.~E.;
  Dendzik,~M.; Michiardi,~M.; Bianchi,~M.; Lizzit,~D.; Presel,~F.
  \latin{et~al.}  Epitaxial Growth of Single-Orientation High-Quality MoS$_2$
  Monolayers. \emph{2D Mater.} \textbf{2018}, \emph{5}, 035012\relax
\mciteBstWouldAddEndPuncttrue
\mciteSetBstMidEndSepPunct{\mcitedefaultmidpunct}
{\mcitedefaultendpunct}{\mcitedefaultseppunct}\relax
\EndOfBibitem
\bibitem[Bruix \latin{et~al.}(2016)Bruix, Miwa, Hauptmann, Wegner, Ulstrup,
  Gr{\o}nborg, Sanders, Dendzik, {\v{C}}abo, Bianchi, Lauritsen, Khajetoorians,
  Hammer, and Hofmann]{Bruix2016}
Bruix,~A.; Miwa,~J.~A.; Hauptmann,~N.; Wegner,~D.; Ulstrup,~S.;
  Gr{\o}nborg,~S.~S.; Sanders,~C.~E.; Dendzik,~M.; {\v{C}}abo,~A.~G.;
  Bianchi,~M. \latin{et~al.}  Single-Layer MoS$_2$ on Au(111): Band Gap
  Renormalization and Substrate Interaction. \emph{Phys. Rev. B} \textbf{2016},
  \emph{93}, 165422\relax
\mciteBstWouldAddEndPuncttrue
\mciteSetBstMidEndSepPunct{\mcitedefaultmidpunct}
{\mcitedefaultendpunct}{\mcitedefaultseppunct}\relax
\EndOfBibitem
\bibitem[Krane \latin{et~al.}(2018)Krane, Lotze, and Franke]{Krane2018}
Krane,~N.; Lotze,~C.; Franke,~K.~J. Moir{\'e} Structure of MoS$_2$ on Au(111):
  Local Structural and Electronic Properties. \emph{Surf. Sci.} \textbf{2018},
  \emph{678}, 136--142\relax
\mciteBstWouldAddEndPuncttrue
\mciteSetBstMidEndSepPunct{\mcitedefaultmidpunct}
{\mcitedefaultendpunct}{\mcitedefaultseppunct}\relax
\EndOfBibitem
\bibitem[Krane \latin{et~al.}(2016)Krane, Lotze, L\"{a}ger, Reecht, and
  Franke]{Krane2016}
Krane,~N.; Lotze,~C.; L\"{a}ger,~J.~M.; Reecht,~G.; Franke,~K.~J. Electronic
  Structure and Luminescence of Quasi-Freestanding MoS$_2$ Nanopatches on
  Au(111). \emph{Nano Lett.} \textbf{2016}, \emph{16}, 5163--5168\relax
\mciteBstWouldAddEndPuncttrue
\mciteSetBstMidEndSepPunct{\mcitedefaultmidpunct}
{\mcitedefaultendpunct}{\mcitedefaultseppunct}\relax
\EndOfBibitem
\bibitem[Varykhalov \latin{et~al.}(2008)Varykhalov, S{\'a}nchez-Barriga,
  Shikin, Biswas, Vescovo, Rybkin, Marchenko, and Rader]{Varykhalov2008}
Varykhalov,~A.; S{\'a}nchez-Barriga,~J.; Shikin,~A.; Biswas,~C.; Vescovo,~E.;
  Rybkin,~A.; Marchenko,~D.; Rader,~O. Electronic and Magnetic Properties of
  Quasifreestanding Graphene on Ni. \emph{Phys. Rev. Lett.} \textbf{2008},
  \emph{101}, 157601\relax
\mciteBstWouldAddEndPuncttrue
\mciteSetBstMidEndSepPunct{\mcitedefaultmidpunct}
{\mcitedefaultendpunct}{\mcitedefaultseppunct}\relax
\EndOfBibitem
\bibitem[Riedl \latin{et~al.}(2009)Riedl, Coletti, Iwasaki, Zakharov, and
  Starke]{Riedl2009}
Riedl,~C.; Coletti,~C.; Iwasaki,~T.; Zakharov,~A.; Starke,~U.
  Quasi-Free-Standing Epitaxial Graphene on SiC Obtained by Hydrogen
  Intercalation. \emph{Phys. Rev. Lett.} \textbf{2009}, \emph{103},
  246804\relax
\mciteBstWouldAddEndPuncttrue
\mciteSetBstMidEndSepPunct{\mcitedefaultmidpunct}
{\mcitedefaultendpunct}{\mcitedefaultseppunct}\relax
\EndOfBibitem
\bibitem[Mahatha \latin{et~al.}(2018)Mahatha, Dendzik, Sanders, Michiardi,
  Bianchi, Miwa, and Hofmann]{Mahatha2018}
Mahatha,~S.~K.; Dendzik,~M.; Sanders,~C.~E.; Michiardi,~M.; Bianchi,~M.;
  Miwa,~J.~A.; Hofmann,~P. Quasi-Free-Standing Single-Layer WS$_2$ Achieved by
  Intercalation. \emph{Phys. Rev. Materials} \textbf{2018}, \emph{2},
  124001\relax
\mciteBstWouldAddEndPuncttrue
\mciteSetBstMidEndSepPunct{\mcitedefaultmidpunct}
{\mcitedefaultendpunct}{\mcitedefaultseppunct}\relax
\EndOfBibitem
\bibitem[Friend and Yoffe(1987)Friend, and Yoffe]{Friend1987}
Friend,~R.; Yoffe,~A. Electronic Properties of Intercalation Complexes of the
  Transition Metal Dichalcogenides. \emph{Adv. Phys.} \textbf{1987}, \emph{36},
  1--94\relax
\mciteBstWouldAddEndPuncttrue
\mciteSetBstMidEndSepPunct{\mcitedefaultmidpunct}
{\mcitedefaultendpunct}{\mcitedefaultseppunct}\relax
\EndOfBibitem
\bibitem[Murugan \latin{et~al.}(2006)Murugan, Quintin, Delville, Campet,
  Gopinath, and Vijayamohanan]{Murugan2006}
Murugan,~A.~V.; Quintin,~M.; Delville,~M.-H.; Campet,~G.; Gopinath,~C.~S.;
  Vijayamohanan,~K. Exfoliation-Induced Nanoribbon Formation of
  Poly(3,4-Ethylene Dioxythiophene) PEDOT Between MoS$_2$ Layers as Cathode
  Material for Lithium Batteries. \emph{J. Power Sources} \textbf{2006},
  \emph{156}, 615--619\relax
\mciteBstWouldAddEndPuncttrue
\mciteSetBstMidEndSepPunct{\mcitedefaultmidpunct}
{\mcitedefaultendpunct}{\mcitedefaultseppunct}\relax
\EndOfBibitem
\bibitem[Whittingham(1976)]{Whittingham1976}
Whittingham,~M.~S. Electrical Energy Storage and Intercalation Chemistry.
  \emph{Science} \textbf{1976}, \emph{192}, 1126--1127\relax
\mciteBstWouldAddEndPuncttrue
\mciteSetBstMidEndSepPunct{\mcitedefaultmidpunct}
{\mcitedefaultendpunct}{\mcitedefaultseppunct}\relax
\EndOfBibitem
\bibitem[Feng \latin{et~al.}(2009)Feng, Ma, Li, Zeng, Guo, and Liu]{Feng2009}
Feng,~C.; Ma,~J.; Li,~H.; Zeng,~R.; Guo,~Z.; Liu,~H. Synthesis of Molybdenum
  Disulfide (MoS$_2$) for Lithium Ion Battery Applications. \emph{Mater. Res.
  Bull.} \textbf{2009}, \emph{44}, 1811--1815\relax
\mciteBstWouldAddEndPuncttrue
\mciteSetBstMidEndSepPunct{\mcitedefaultmidpunct}
{\mcitedefaultendpunct}{\mcitedefaultseppunct}\relax
\EndOfBibitem
\bibitem[Li \latin{et~al.}(2019)Li, Jiang, Khan, Goswami, Liu, Passian, and
  Thundat]{Li2019}
Li,~Z.; Jiang,~K.; Khan,~F.; Goswami,~A.; Liu,~J.; Passian,~A.; Thundat,~T.
  Anomalous Interfacial Stress Generation During Sodium
  Intercalation/Extraction in MoS$_2$ Thin-Film Anodes. \emph{Sci. Adv.}
  \textbf{2019}, \emph{5}, eaav2820\relax
\mciteBstWouldAddEndPuncttrue
\mciteSetBstMidEndSepPunct{\mcitedefaultmidpunct}
{\mcitedefaultendpunct}{\mcitedefaultseppunct}\relax
\EndOfBibitem
\bibitem[Py and Haering(1983)Py, and Haering]{Py1983}
Py,~M.; Haering,~R. Structural Destabilization Induced by Lithium Intercalation
  in MoS$_2$ and Related Compounds. \emph{Canadian J. Phys.} \textbf{1983},
  \emph{61}, 76--84\relax
\mciteBstWouldAddEndPuncttrue
\mciteSetBstMidEndSepPunct{\mcitedefaultmidpunct}
{\mcitedefaultendpunct}{\mcitedefaultseppunct}\relax
\EndOfBibitem
\bibitem[Wypych and Sch{\"o}llhorn(1992)Wypych, and Sch{\"o}llhorn]{Wypych1992}
Wypych,~F.; Sch{\"o}llhorn,~R. 1T-MoS$_2$, a New Metallic Modification of
  Molybdenum Disulfide. \emph{J. Chem. Soc. Chem. Comm.} \textbf{1992},
  1386--1388\relax
\mciteBstWouldAddEndPuncttrue
\mciteSetBstMidEndSepPunct{\mcitedefaultmidpunct}
{\mcitedefaultendpunct}{\mcitedefaultseppunct}\relax
\EndOfBibitem
\bibitem[Heising and Kanatzidis(1999)Heising, and Kanatzidis]{Heising1999}
Heising,~J.; Kanatzidis,~M.~G. Structure of Restacked MoS$_2$ and WS$_2$
  Elucidated by Electron Crystallography. \emph{J. Am. Chem. Soc.}
  \textbf{1999}, \emph{121}, 638--643\relax
\mciteBstWouldAddEndPuncttrue
\mciteSetBstMidEndSepPunct{\mcitedefaultmidpunct}
{\mcitedefaultendpunct}{\mcitedefaultseppunct}\relax
\EndOfBibitem
\bibitem[Eda \latin{et~al.}(2011)Eda, Yamaguchi, Voiry, Fujita, Chen, and
  Chhowalla]{Eda2011}
Eda,~G.; Yamaguchi,~H.; Voiry,~D.; Fujita,~T.; Chen,~M.; Chhowalla,~M.
  Photoluminescence from Chemically Exfoliated MoS$_2$. \emph{Nano Lett.}
  \textbf{2011}, \emph{11}, 5111--5116\relax
\mciteBstWouldAddEndPuncttrue
\mciteSetBstMidEndSepPunct{\mcitedefaultmidpunct}
{\mcitedefaultendpunct}{\mcitedefaultseppunct}\relax
\EndOfBibitem
\bibitem[Wang \latin{et~al.}(2013)Wang, Ou, Balendhran, Chrimes, Mortazavi,
  Yao, Field, Latham, Bansal, Friend, Zhuiykov, Medhekar, and
  Kalantar-zadeh]{Wang2013}
Wang,~Y.; Ou,~J.~Z.; Balendhran,~S.; Chrimes,~A.~F.; Mortazavi,~M.; Yao,~D.~D.;
  Field,~M.~R.; Latham,~K.; Bansal,~V.; Friend,~J.~R. \latin{et~al.}
  Electrochemical Control of Photoluminescence in Two-Dimensional MoS$_2$
  Nanoflakes. \emph{ACS Nano} \textbf{2013}, \emph{7}, 10083--10093\relax
\mciteBstWouldAddEndPuncttrue
\mciteSetBstMidEndSepPunct{\mcitedefaultmidpunct}
{\mcitedefaultendpunct}{\mcitedefaultseppunct}\relax
\EndOfBibitem
\bibitem[Wang \latin{et~al.}(2014)Wang, Shen, Wang, Yu, and Chen]{Wang2014}
Wang,~X.; Shen,~X.; Wang,~Z.; Yu,~R.; Chen,~L. Atomic-Scale Clarification of
  Structural Transition of MoS$_2$ Upon Sodium Intercalation. \emph{ACS Nano}
  \textbf{2014}, \emph{8}, 11394--11400\relax
\mciteBstWouldAddEndPuncttrue
\mciteSetBstMidEndSepPunct{\mcitedefaultmidpunct}
{\mcitedefaultendpunct}{\mcitedefaultseppunct}\relax
\EndOfBibitem
\bibitem[Fan \latin{et~al.}(2015)Fan, Xu, Zhou, Sun, Li, Nguyen, Terrones, and
  Mallouk]{Fan2015}
Fan,~X.; Xu,~P.; Zhou,~D.; Sun,~Y.; Li,~Y.~C.; Nguyen,~M. A.~T.; Terrones,~M.;
  Mallouk,~T.~E. Fast and Efficient Preparation of Exfoliated 2H-MoS$_2$
  Nanosheets by Sonication-Assisted Lithium Intercalation and Infrared
  Laser-Induced 1T to 2H Phase Reversion. \emph{Nano Lett.} \textbf{2015},
  \emph{15}, 5956--5960\relax
\mciteBstWouldAddEndPuncttrue
\mciteSetBstMidEndSepPunct{\mcitedefaultmidpunct}
{\mcitedefaultendpunct}{\mcitedefaultseppunct}\relax
\EndOfBibitem
\bibitem[Guo \latin{et~al.}(2015)Guo, Sun, Ouyang, Raja, Song, Heinz, and
  Brus]{Guo2015}
Guo,~Y.; Sun,~D.; Ouyang,~B.; Raja,~A.; Song,~J.; Heinz,~T.~F.; Brus,~L.~E.
  Probing the Dynamics of the Metallic-to-Semiconducting Structural Phase
  Transformation in MoS$_2$ Crystals. \emph{Nano Lett.} \textbf{2015},
  \emph{15}, 5081--5088\relax
\mciteBstWouldAddEndPuncttrue
\mciteSetBstMidEndSepPunct{\mcitedefaultmidpunct}
{\mcitedefaultendpunct}{\mcitedefaultseppunct}\relax
\EndOfBibitem
\bibitem[Xiong \latin{et~al.}(2015)Xiong, Wang, Liu, Sun, Brongersma, Pop, and
  Cui]{Xiong2015}
Xiong,~F.; Wang,~H.; Liu,~X.; Sun,~J.; Brongersma,~M.; Pop,~E.; Cui,~Y. Li
  Intercalation in MoS$_2$: In Situ Observation of its Dynamics and Tuning
  Optical and Electrical Properties. \emph{Nano Lett.} \textbf{2015},
  \emph{15}, 6777--6784\relax
\mciteBstWouldAddEndPuncttrue
\mciteSetBstMidEndSepPunct{\mcitedefaultmidpunct}
{\mcitedefaultendpunct}{\mcitedefaultseppunct}\relax
\EndOfBibitem
\bibitem[Ahmad \latin{et~al.}(2017)Ahmad, M{\"u}ller, Habenicht, Schuster,
  Knupfer, and B{\"u}chner]{Ahmad2017}
Ahmad,~M.; M{\"u}ller,~E.; Habenicht,~C.; Schuster,~R.; Knupfer,~M.;
  B{\"u}chner,~B. Semiconductor-to-Metal Transition in the Bulk of WSe$_2$ Upon
  Potassium Intercalation. \emph{J Physics: Condens. Matter.} \textbf{2017},
  \emph{29}, 165502\relax
\mciteBstWouldAddEndPuncttrue
\mciteSetBstMidEndSepPunct{\mcitedefaultmidpunct}
{\mcitedefaultendpunct}{\mcitedefaultseppunct}\relax
\EndOfBibitem
\bibitem[Hoffmann \latin{et~al.}(2004)Hoffmann, S{\o}ndergaard, Schultz, Li,
  and Hofmann]{Hoffmann2004}
Hoffmann,~S.; S{\o}ndergaard,~C.; Schultz,~C.; Li,~Z.; Hofmann,~P. An
  Undulator-Based Spherical Grating Monochromator Beamline for Angle-Resolved
  Photoemission Spectroscopy. \emph{Nucl. Instrum. Methods Phys. Res. Sect. A}
  \textbf{2004}, \emph{523}, 441--453\relax
\mciteBstWouldAddEndPuncttrue
\mciteSetBstMidEndSepPunct{\mcitedefaultmidpunct}
{\mcitedefaultendpunct}{\mcitedefaultseppunct}\relax
\EndOfBibitem
\bibitem[Taborek and Rutledge(1992)Taborek, and Rutledge]{Taborek1992}
Taborek,~P.; Rutledge,~J. Novel Wetting Behavior of $^4$He on Cesium.
  \emph{Phys. Rev. Lett.} \textbf{1992}, \emph{68}, 2184\relax
\mciteBstWouldAddEndPuncttrue
\mciteSetBstMidEndSepPunct{\mcitedefaultmidpunct}
{\mcitedefaultendpunct}{\mcitedefaultseppunct}\relax
\EndOfBibitem
\bibitem[Petrovi{\'c} \latin{et~al.}(2013)Petrovi{\'c}, Raki{\'c}, Runte,
  Busse, Sadowski, Lazi{\'c}, Pletikosi{\'c}, Pan, Milun, Pervan, Atodiresei,
  Brako, \v{S}ok\v{c}evi\'{c}, Valla, Michely, and Kralj]{Petrovic2013}
Petrovi{\'c},~M.; Raki{\'c},~I.~{\v{S}}.; Runte,~S.; Busse,~C.; Sadowski,~J.;
  Lazi{\'c},~P.; Pletikosi{\'c},~I.; Pan,~Z.-H.; Milun,~M.; Pervan,~P.
  \latin{et~al.}  The Mechanism of Caesium Intercalation of Graphene.
  \emph{Nat. Commun.} \textbf{2013}, \emph{4}, 1--8\relax
\mciteBstWouldAddEndPuncttrue
\mciteSetBstMidEndSepPunct{\mcitedefaultmidpunct}
{\mcitedefaultendpunct}{\mcitedefaultseppunct}\relax
\EndOfBibitem
\bibitem[Vlieg(2000)]{Vlieg2000}
Vlieg,~E. ROD: a Program for Surface X-Ray Crystallography. \emph{J. Appl.
  Cryst.} \textbf{2000}, \emph{33}, 401--405\relax
\mciteBstWouldAddEndPuncttrue
\mciteSetBstMidEndSepPunct{\mcitedefaultmidpunct}
{\mcitedefaultendpunct}{\mcitedefaultseppunct}\relax
\EndOfBibitem
\bibitem[Lewis \latin{et~al.}(2011)Lewis, Jel{\'\i}nek, Ortega, Demkov,
  Trabada, Haycock, Wang, Adams, Tomfohr, Abad, Wang, and Drabold]{Lewis2011}
Lewis,~J.~P.; Jel{\'\i}nek,~P.; Ortega,~J.; Demkov,~A.~A.; Trabada,~D.~G.;
  Haycock,~B.; Wang,~H.; Adams,~G.; Tomfohr,~J.~K.; Abad,~E. \latin{et~al.}
  Advances and Applications in the FIREBALL Ab Initio Tight-Binding
  Molecular-Dynamics Formalism. \emph{Phys. Stat. Sol. b} \textbf{2011},
  \emph{248}, 1989--2007\relax
\mciteBstWouldAddEndPuncttrue
\mciteSetBstMidEndSepPunct{\mcitedefaultmidpunct}
{\mcitedefaultendpunct}{\mcitedefaultseppunct}\relax
\EndOfBibitem
\bibitem[Jel{\'\i}nek \latin{et~al.}(2005)Jel{\'\i}nek, Wang, Lewis, Sankey,
  and Ortega]{Jelinek2005}
Jel{\'\i}nek,~P.; Wang,~H.; Lewis,~J.~P.; Sankey,~O.~F.; Ortega,~J. Multicenter
  Approach to the Exchange-Correlation Interactions in Ab Initio Tight-Binding
  Methods. \emph{Phys. Rev. B} \textbf{2005}, \emph{71}, 235101\relax
\mciteBstWouldAddEndPuncttrue
\mciteSetBstMidEndSepPunct{\mcitedefaultmidpunct}
{\mcitedefaultendpunct}{\mcitedefaultseppunct}\relax
\EndOfBibitem
\bibitem[Sankey and Niklewski(1989)Sankey, and Niklewski]{Sankey1989}
Sankey,~O.~F.; Niklewski,~D.~J. Ab Initio Multicenter Tight-Binding Model for
  Molecular-Dynamics Simulations and Other Applications in Covalent Systems.
  \emph{Phys. Rev. B} \textbf{1989}, \emph{40}, 3979\relax
\mciteBstWouldAddEndPuncttrue
\mciteSetBstMidEndSepPunct{\mcitedefaultmidpunct}
{\mcitedefaultendpunct}{\mcitedefaultseppunct}\relax
\EndOfBibitem
\bibitem[Gonz{\'a}lez \latin{et~al.}(2016)Gonz{\'a}lez, Biel, and
  Dappe]{Gonzalez2016}
Gonz{\'a}lez,~C.; Biel,~B.; Dappe,~Y.~J. Theoretical Characterisation of Point
  Defects on a {MoS$_2$} Monolayer by Scanning Tunnelling Microscopy.
  \emph{Nanotechnology} \textbf{2016}, \emph{27}, 105702\relax
\mciteBstWouldAddEndPuncttrue
\mciteSetBstMidEndSepPunct{\mcitedefaultmidpunct}
{\mcitedefaultendpunct}{\mcitedefaultseppunct}\relax
\EndOfBibitem
\bibitem[Gr{\o}nborg \latin{et~al.}(2015)Gr{\o}nborg, Ulstrup, Bianchi,
  Dendzik, Sanders, Lauritsen, Hofmann, and Miwa]{Gronborg2015}
Gr{\o}nborg,~S.~S.; Ulstrup,~S.; Bianchi,~M.; Dendzik,~M.; Sanders,~C.~E.;
  Lauritsen,~J.~V.; Hofmann,~P.; Miwa,~J.~A. Synthesis of Epitaxial
  Single-Layer MoS$_2$ on Au(111). \emph{Langmuir} \textbf{2015}, \emph{31},
  9700--9706\relax
\mciteBstWouldAddEndPuncttrue
\mciteSetBstMidEndSepPunct{\mcitedefaultmidpunct}
{\mcitedefaultendpunct}{\mcitedefaultseppunct}\relax
\EndOfBibitem
\bibitem[Bollinger \latin{et~al.}(2001)Bollinger, Lauritsen, Jacobsen,
  N{\o}rskov, Helveg, and Besenbacher]{Bollinger2000}
Bollinger,~M.; Lauritsen,~J.; Jacobsen,~K.~W.; N{\o}rskov,~J.~K.; Helveg,~S.;
  Besenbacher,~F. One-Dimensional Metallic Edge States in MoS$_2$. \emph{Phys.
  Rev. Lett.} \textbf{2001}, \emph{87}, 196803\relax
\mciteBstWouldAddEndPuncttrue
\mciteSetBstMidEndSepPunct{\mcitedefaultmidpunct}
{\mcitedefaultendpunct}{\mcitedefaultseppunct}\relax
\EndOfBibitem
\bibitem[Guinier(1994)]{Guinier1994}
Guinier,~A. \emph{{X}-ray diffraction in crystals, imperfect crystals, and
  amorphous bodies}; Dover Publications, 1994\relax
\mciteBstWouldAddEndPuncttrue
\mciteSetBstMidEndSepPunct{\mcitedefaultmidpunct}
{\mcitedefaultendpunct}{\mcitedefaultseppunct}\relax
\EndOfBibitem
\bibitem[El-Mahalawy and Evans(1976)El-Mahalawy, and Evans]{ElMahalawy1976}
El-Mahalawy,~S.; Evans,~B. The Thermal Expansion of 2H-MoS$_2$, 2H-MoSe$_2$ and
  2H-WSe$_2$ between 20 and 800$^\circ$C. \emph{J. Appl. Cryst.} \textbf{1976},
  \emph{9}, 403--406\relax
\mciteBstWouldAddEndPuncttrue
\mciteSetBstMidEndSepPunct{\mcitedefaultmidpunct}
{\mcitedefaultendpunct}{\mcitedefaultseppunct}\relax
\EndOfBibitem
\bibitem[Blanc \latin{et~al.}(2012)Blanc, Coraux, Vo-Van, N'Diaye, Geaymond,
  and Renaud]{Blanc2012}
Blanc,~N.; Coraux,~J.; Vo-Van,~C.; N'Diaye,~A.~T.; Geaymond,~O.; Renaud,~G.
  {Local Deformations and Incommensurability of High-Quality Epitaxial Graphene
  on a Weakly Interacting Transition Metal}. \emph{Phys. Rev. B} \textbf{2012},
  \emph{86}, 235439\relax
\mciteBstWouldAddEndPuncttrue
\mciteSetBstMidEndSepPunct{\mcitedefaultmidpunct}
{\mcitedefaultendpunct}{\mcitedefaultseppunct}\relax
\EndOfBibitem
\bibitem[Martoccia \latin{et~al.}(2010)Martoccia, Bj{\"o}rck, Schlep{\"u}tz,
  Brugger, Pauli, Patterson, Greber, and Willmott]{Martoccia2010}
Martoccia,~D.; Bj{\"o}rck,~M.; Schlep{\"u}tz,~C.; Brugger,~T.; Pauli,~S.;
  Patterson,~B.; Greber,~T.; Willmott,~P. {Graphene on Ru(0001): a Corrugated
  and Chiral Structure}. \emph{New J. Phys.} \textbf{2010}, \emph{12},
  043028\relax
\mciteBstWouldAddEndPuncttrue
\mciteSetBstMidEndSepPunct{\mcitedefaultmidpunct}
{\mcitedefaultendpunct}{\mcitedefaultseppunct}\relax
\EndOfBibitem
\bibitem[Martoccia \latin{et~al.}(2008)Martoccia, Willmott, Brugger,
  Bj{\"o}rck, G{\"u}nther, Schlep{\"u}tz, Cervellino, Pauli, Patterson,
  Marchini, Wintterlin, Moritz, and Greber]{Martoccia2008}
Martoccia,~D.; Willmott,~P.; Brugger,~T.; Bj{\"o}rck,~M.; G{\"u}nther,~S.;
  Schlep{\"u}tz,~C.; Cervellino,~A.; Pauli,~S.; Patterson,~B.; Marchini,~S.
  \latin{et~al.}  {Graphene on Ru (0001): a 25$\times$ 25 Supercell}.
  \emph{Phys. Rev. Lett.} \textbf{2008}, \emph{101}, 126102\relax
\mciteBstWouldAddEndPuncttrue
\mciteSetBstMidEndSepPunct{\mcitedefaultmidpunct}
{\mcitedefaultendpunct}{\mcitedefaultseppunct}\relax
\EndOfBibitem
\bibitem[Robinson(1986)]{Robinson1986}
Robinson,~I.~K. Crystal Truncation Rods and Surface Roughness. \emph{Phys. Rev.
  B} \textbf{1986}, \emph{33}, 3830\relax
\mciteBstWouldAddEndPuncttrue
\mciteSetBstMidEndSepPunct{\mcitedefaultmidpunct}
{\mcitedefaultendpunct}{\mcitedefaultseppunct}\relax
\EndOfBibitem
\bibitem[Warmuth \latin{et~al.}(2016)Warmuth, Bruix, Michiardi, H{\"a}nke,
  Bianchi, Wiebe, Wiesendanger, Hammer, Hofmann, and
  Khajetoorians]{Warmuth2016}
Warmuth,~J.; Bruix,~A.; Michiardi,~M.; H{\"a}nke,~T.; Bianchi,~M.; Wiebe,~J.;
  Wiesendanger,~R.; Hammer,~B.; Hofmann,~P.; Khajetoorians,~A.~A. Band-Gap
  Engineering by Bi Intercalation of Graphene on Ir(111). \emph{Phys. Rev. B}
  \textbf{2016}, \emph{93}, 165437\relax
\mciteBstWouldAddEndPuncttrue
\mciteSetBstMidEndSepPunct{\mcitedefaultmidpunct}
{\mcitedefaultendpunct}{\mcitedefaultseppunct}\relax
\EndOfBibitem
\bibitem[Huang \latin{et~al.}(2014)Huang, Gong, and Zeng]{Huang2014}
Huang,~L.~F.; Gong,~P.~L.; Zeng,~Z. Correlation between Structure, Phonon
  Spectra, Thermal Expansion, and Thermomechanics of Single-Layer MoS$_2$.
  \emph{Phys. Rev. B} \textbf{2014}, \emph{90}, 045409\relax
\mciteBstWouldAddEndPuncttrue
\mciteSetBstMidEndSepPunct{\mcitedefaultmidpunct}
{\mcitedefaultendpunct}{\mcitedefaultseppunct}\relax
\EndOfBibitem
\bibitem[Nix and MacNair(1941)Nix, and MacNair]{Nix1941}
Nix,~F.; MacNair,~D. The Thermal Expansion of Pure Metals: Copper, Gold,
  Aluminum, Nickel, and Iron. \emph{Phys. Rev.} \textbf{1941}, \emph{60},
  597\relax
\mciteBstWouldAddEndPuncttrue
\mciteSetBstMidEndSepPunct{\mcitedefaultmidpunct}
{\mcitedefaultendpunct}{\mcitedefaultseppunct}\relax
\EndOfBibitem
\bibitem[Brumme \latin{et~al.}(2015)Brumme, Calandra, and Mauri]{Brumme2015}
Brumme,~T.; Calandra,~M.; Mauri,~F. First-Principles Theory of Field-Effect
  Doping in Transition-Metal Dichalcogenides: Structural Properties, Electronic
  Structure, Hall Coefficient, and Electrical Conductivity. \emph{Phys. Rev. B}
  \textbf{2015}, \emph{91}, 155436\relax
\mciteBstWouldAddEndPuncttrue
\mciteSetBstMidEndSepPunct{\mcitedefaultmidpunct}
{\mcitedefaultendpunct}{\mcitedefaultseppunct}\relax
\EndOfBibitem
\bibitem[Duerloo \latin{et~al.}(2014)Duerloo, Li, and Reed]{Duerloo2014}
Duerloo,~K.-A.~N.; Li,~Y.; Reed,~E.~J. Structural Phase Transitions in
  Two-Dimensional Mo-and W-Dichalcogenide Monolayers. \emph{Nat. Commun.}
  \textbf{2014}, \emph{5}, 4214\relax
\mciteBstWouldAddEndPuncttrue
\mciteSetBstMidEndSepPunct{\mcitedefaultmidpunct}
{\mcitedefaultendpunct}{\mcitedefaultseppunct}\relax
\EndOfBibitem
\bibitem[Tan \latin{et~al.}(2018)Tan, Sarkar, Zhao, Luo, Luo, Poh, Abdelwahab,
  Zhou, Venkatesan, Chen, Quek, and Loh]{Tan2018}
Tan,~S. J.~R.; Sarkar,~S.; Zhao,~X.; Luo,~X.; Luo,~Y.~Z.; Poh,~S.~M.;
  Abdelwahab,~I.; Zhou,~W.; Venkatesan,~T.; Chen,~W. \latin{et~al.}
  Temperature- and Phase-Dependent Phonon Renormalization in 1T'-MoS$_2$.
  \emph{ACS Nano} \textbf{2018}, \emph{12}, 5051--5058\relax
\mciteBstWouldAddEndPuncttrue
\mciteSetBstMidEndSepPunct{\mcitedefaultmidpunct}
{\mcitedefaultendpunct}{\mcitedefaultseppunct}\relax
\EndOfBibitem
\bibitem[Andersen \latin{et~al.}(2012)Andersen, Kathmann, Lilga, Albrecht,
  Hallen, and Mei]{Andersen2012}
Andersen,~A.; Kathmann,~S.~M.; Lilga,~M.~A.; Albrecht,~K.~O.; Hallen,~R.~T.;
  Mei,~D. First-Principles Characterization of Potassium Intercalation in
  Hexagonal 2H-MoS$_2$. \emph{J. Phys. Chem. C} \textbf{2012}, \emph{116},
  1826--1832\relax
\mciteBstWouldAddEndPuncttrue
\mciteSetBstMidEndSepPunct{\mcitedefaultmidpunct}
{\mcitedefaultendpunct}{\mcitedefaultseppunct}\relax
\EndOfBibitem
\bibitem[Mart{\'\i}nez-Galera \latin{et~al.}(2016)Mart{\'\i}nez-Galera,
  Schr{\"o}der, Huttmann, Jolie, Craes, Busse, Caciuc, Atodiresei, Bl{\"u}gel,
  and Michely]{Martinez2016}
Mart{\'\i}nez-Galera,~A.~J.; Schr{\"o}der,~U.; Huttmann,~F.; Jolie,~W.;
  Craes,~F.; Busse,~C.; Caciuc,~V.; Atodiresei,~N.; Bl{\"u}gel,~S.; Michely,~T.
  Oxygen Orders Differently Under Graphene: New Superstructures on Ir (111).
  \emph{Nanoscale} \textbf{2016}, \emph{8}, 1932--1943\relax
\mciteBstWouldAddEndPuncttrue
\mciteSetBstMidEndSepPunct{\mcitedefaultmidpunct}
{\mcitedefaultendpunct}{\mcitedefaultseppunct}\relax
\EndOfBibitem
\bibitem[Somoano \latin{et~al.}(1973)Somoano, Hadek, and Rembaum]{Somoano1973}
Somoano,~R.; Hadek,~V.; Rembaum,~A. Alkali Metal Intercalates of Molybdenum
  Disulfide. \emph{J. Chem. Phys.} \textbf{1973}, \emph{58}, 697--701\relax
\mciteBstWouldAddEndPuncttrue
\mciteSetBstMidEndSepPunct{\mcitedefaultmidpunct}
{\mcitedefaultendpunct}{\mcitedefaultseppunct}\relax
\EndOfBibitem
\bibitem[Lindgren and Walld{\'e}n(1980)Lindgren, and Walld{\'e}n]{Lindgren1980}
Lindgren,~S.; Walld{\'e}n,~L. {Photoemission of Electrons at the Cu(111)/Na
  Interface}. \emph{Solid State Commun.} \textbf{1980}, \emph{34},
  671--673\relax
\mciteBstWouldAddEndPuncttrue
\mciteSetBstMidEndSepPunct{\mcitedefaultmidpunct}
{\mcitedefaultendpunct}{\mcitedefaultseppunct}\relax
\EndOfBibitem
\bibitem[Forster \latin{et~al.}(2006)Forster, Bendounan, Ziroff, and
  Reinert]{Forster2006}
Forster,~F.; Bendounan,~A.; Ziroff,~J.; Reinert,~F. {Systematic Studies on
  Surface Modifications by ARUPS on Shockley-Type Surface States}. \emph{Surf.
  Sci.} \textbf{2006}, \emph{600}, 3870--3874\relax
\mciteBstWouldAddEndPuncttrue
\mciteSetBstMidEndSepPunct{\mcitedefaultmidpunct}
{\mcitedefaultendpunct}{\mcitedefaultseppunct}\relax
\EndOfBibitem
\bibitem[Miwa \latin{et~al.}(2015)Miwa, Ulstrup, S{\o}rensen, Dendzik,
  {\v{C}}abo, Bianchi, Lauritsen, and Hofmann]{Miwa2015}
Miwa,~J.~A.; Ulstrup,~S.; S{\o}rensen,~S.~G.; Dendzik,~M.; {\v{C}}abo,~A.~G.;
  Bianchi,~M.; Lauritsen,~J.~V.; Hofmann,~P. Electronic Structure of Epitaxial
  Single-Layer MoS$_2$. \emph{Phys. Rev. Lett.} \textbf{2015}, \emph{114},
  046802\relax
\mciteBstWouldAddEndPuncttrue
\mciteSetBstMidEndSepPunct{\mcitedefaultmidpunct}
{\mcitedefaultendpunct}{\mcitedefaultseppunct}\relax
\EndOfBibitem
\end{mcitethebibliography}

\providecommand{\latin}[1]{#1}
\makeatletter
\providecommand{\doi}
  {\begingroup\let\do\@makeother\dospecials
  \catcode`\{=1 \catcode`\}=2\doi@aux}
\providecommand{\doi@aux}[1]{\endgroup\texttt{#1}}
\makeatother
\providecommand*\mcitethebibliography{\thebibliography}
\csname @ifundefined\endcsname{endmcitethebibliography}
  {\let\endmcitethebibliography\endthebibliography}{}

\end{document}